\documentclass[multphys,vecphys]{svmult}

\usepackage{makeidx}     % allows index generation
\usepackage{graphicx}    % standard LaTeX graphics tool
\usepackage{multicol}    % used for the two-column index
\usepackage{amssymb}     % AMS symbols
\makeindex             % used for the subject index
\newcommand{\bbox}[1]{%
     {{\hbox{\boldmath$\displaystyle#1$}}}}
\newcommand{\beq}{\begin{equation}}
\newcommand{\beqa}{\begin{eqnarray}}
\newcommand{\densx}{n({\mathbf x})}
\newcommand{\ds}{\displaystyle}
\newcommand{\eeq}{\end{equation}}
\newcommand{\eeqa}{\end{eqnarray}}

\newcommand{\Exc}{E_{{\rm xc}}}
\newcommand{\FHK}{F_{{\rm HK}}}
\newcommand{\Fni}{F_{{\rm ni}}}
\newcommand{\fpi}{f_\pi}
\newcommand{\GH}{G_H}
\newcommand{\GHinv}{\GH^{-1}}
\newcommand{\grad}{{\bbox{\nabla}}}
\newcommand{\gv}{g_{{\rm v}}}

\newcommand{\kf}{k_{\scriptscriptstyle\rm F}}
\newcommand{\Mstar}{M^*}
\newcommand{\rhoB}{\rho_{{\scriptscriptstyle \rm B}}}
\newcommand{\rhoBt}{\wt\rho_{{\scriptscriptstyle \rm B}}}

\newcommand{\rhost}{\wt\rho_{{\scriptstyle \rm s}}}

\newcommand{\rhothree}{\rho_{3}}
\newcommand{\rhothreet}{\wt\rho_{3}}

\newcommand{\tensor}{{{s}}}
\newcommand{\tensort}{\wt{{s}}}
\newcommand{\Tr}{{\rm Tr\ }}
\newcommand{\wt}{\widetilde}

\begin{document}

\title*{Next Generation Relativistic Models}
%\titlerunning{Short Title}

\author{R.~J.\ Furnstahl}
\institute{Physics Department, The Ohio State University,
  Columbus, OH\ \ 43210
\texttt{furnstahl.1@osu.edu}}
\maketitle

%%%%%%%%%%%%%%%%%%%%%%%%%%%%%%%%%%%%%%%%%%%%%%%%%%%%%%%%%%%%%%%%%%%%%
%%%%%%%%%%%%%%%%%%%%%%%%%%%%%%%%%%%%%%%%%%%%%%%%%%%%%%%%%%%%%%%%%%%%%

%
%\index{paragraph}
% Use the \index{} command to code your index words
%\begin{figure}
%\centering
%\end{figure}
%
%

%%%%%%%%%%%%%%%%%%%%%%%%%%%%%%%%%%%%%%%%%%%%%%%%%%%%%%%%%%%%%%%%%%%%%
%%%%%%%%%%%%%%%%%%%%%%%%%%%%%%%%%%%%%%%%%%%%%%%%%%%%%%%%%%%%%%%%%%%%%

\section{Introduction}
\label{sec:1}

The current generation of covariant mean-field models has had many
successes in calculations of bulk observables for medium to heavy
nuclei, but there remain many open questions \cite{RING96,SEROT97}. 
New challenges are
confronted when trying to  systematically extend these models to
reliably  address nuclear structure physics away from the line of
stability. 
In this lecture, we discuss a framework for  the next
generation of relativistic models that can address these questions and
challenges. 
We interpret nuclear mean-field approaches as approximate
implementations of Kohn-Sham density functional theory (DFT), which is
widely used in condensed matter and quantum chemistry 
applications \cite{KOHN65,DREIZLER90}. 
We
look to effective field theory (EFT) for a systematic approach to
low-energy nuclear physics that can provide the framework for nuclear
DFT \cite{FURNSTAHL00}.

We start with the key principle underlying any effective low-energy
model and then describe how EFT's exploit it to 
systematically remove model dependence from calculations of
low-energy observables.
Chiral EFT for few nucleon systems is maturing rapidly
and serves as a prototype for nuclear structure EFT.
The immediate question is:  why consider a \emph{relativistic} EFT for nuclei?
We show how 
the EFT interpretation of the many-body problem clarifies the role of
the ``Dirac sea'' in relativistic mean-field calculations of
ground states and collective excited states
(relativistic RPA).  
Next, the strengths of EFT-motivated power counting for nuclear
energy functionals is shown
by the analysis of neutron skins in relativistic models,
which also reveals weaknesses in current functionals. 
Finally, we propose 
a formalism for constructing improved
covariant functionals based on EFT/DFT.
We illustrate the basic ideas using 
recent work on deriving systematic
Kohn-Sham functionals for cold atomic gases, which
exhibit the sort of power counting and order-by-order improvement 
in the calculation of observables
that we seek for nuclear functionals.

%%%%%%%%%%%%%%%%%%%%%%%%%%%%%%%%%%%%%%%%%%%%%%%%%%%%%%%%%%%%%%%%%%%%%
%%%%%%%%%%%%%%%%%%%%%%%%%%%%%%%%%%%%%%%%%%%%%%%%%%%%%%%%%%%%%%%%%%%%%

\section{Low-Energy Effective Theories of QCD}
\label{sec:2}

\begin{figure}[t]
  \centering
  \includegraphics*[width=3.5in,angle=0]{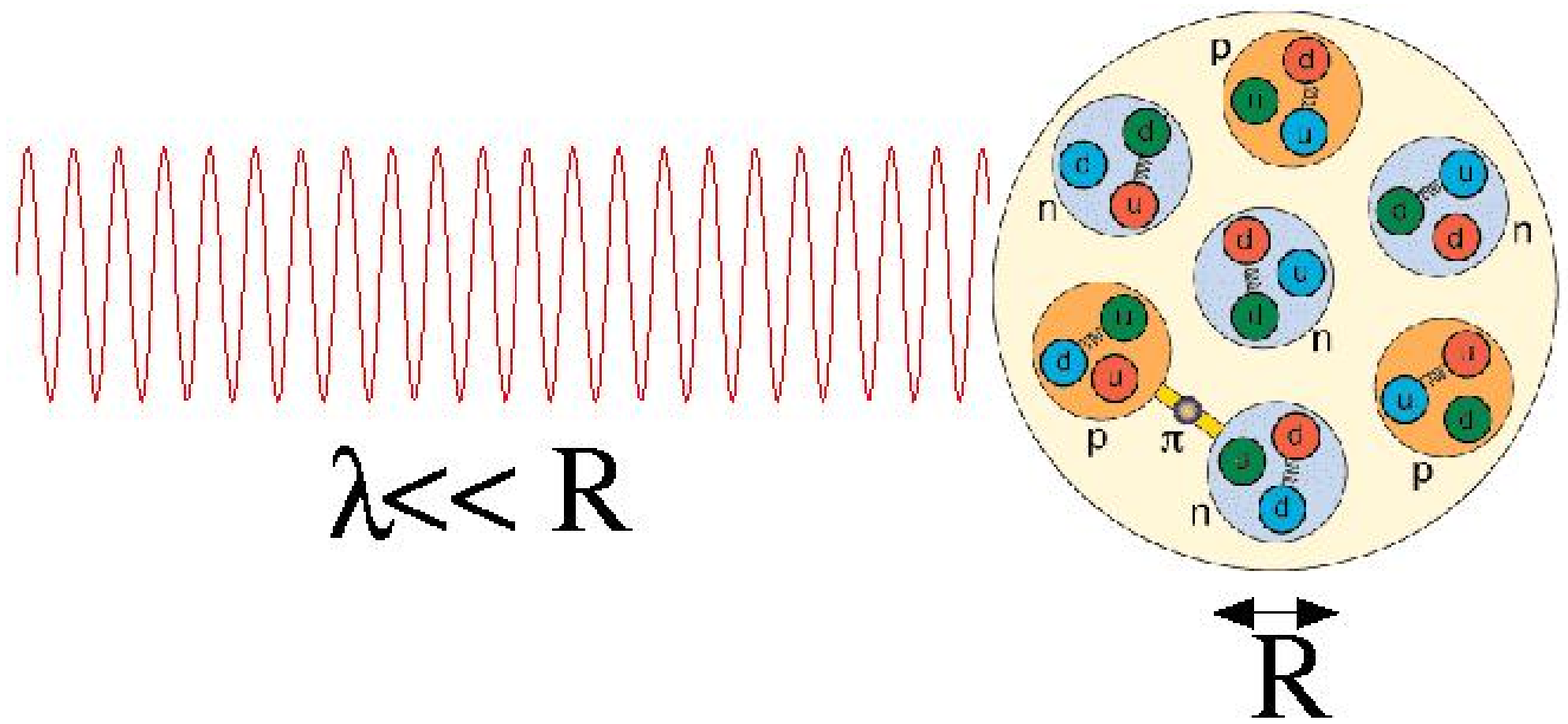}
  \caption{A short wavelength probe resolves 
      quark and gluon degrees of freedom.}
  \label{fig:resol3}       

  \vspace*{.2in}

  \includegraphics*[width=3.5in,angle=0]{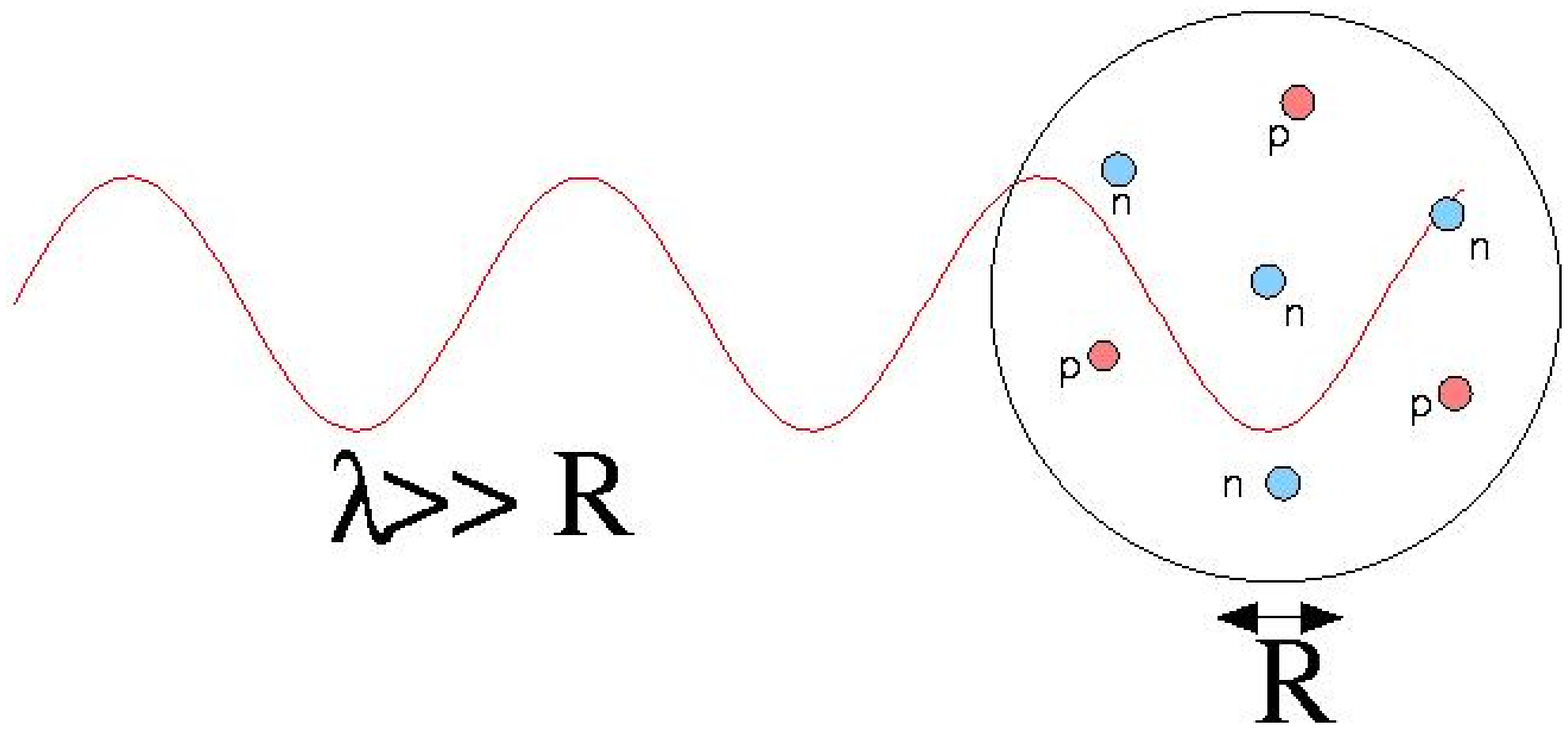}
  \caption{For long wavelength probes, low-energy degrees of freedom
  (protons and neutrons here) 
  with simpler short-distance structure can be used.}
  \label{fig:resol3b}       
\end{figure}

If a system is probed with wavelengths small compared to the size
of characteristic sub-structure,
then details of that substructure are resolved and must be included
explicitly (e.g., see Fig.~\ref{fig:resol3} for a nucleus).
On the other hand, a general principle of \emph{any}
effective low-energy theory is that if a system
is probed or interacts at low energies, resolution is also low, and 
%fine details of 
what happens at short distances or in high-energy
intermediate states is not resolved \cite{LEPAGE,CROSSING}.
In this case, it is easier and more efficient to use low-energy degrees
of freedom for low-energy processes (see Fig.~\ref{fig:resol3b}).
The short-distance structure can be replaced by something simpler
(and wrong at short distances!) without distorting low-energy
observables.
This principle is implicit in conventional 
nonrelativistic nuclear phenomenology with
cut-off nucleon-nucleon (NN) potentials (although rarely acknowledged!).
There are many ways to replace the structure; an illuminating way is
to lower a cutoff $\Lambda$ on intermediate states.

\subsection{Renormalization Group and NN Potential}
\label{subsec:2a}

\begin{figure}[t]
 \centering 
 \includegraphics*[width=2.5in,angle=0]{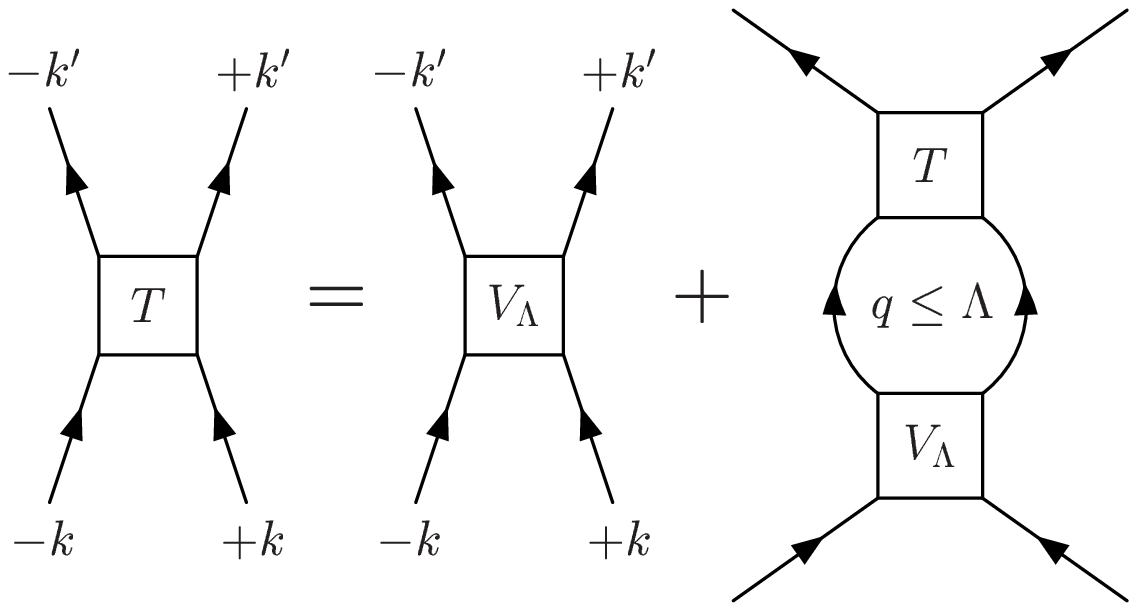}
 \hfill
 \includegraphics*[width=1.7in,angle=0]{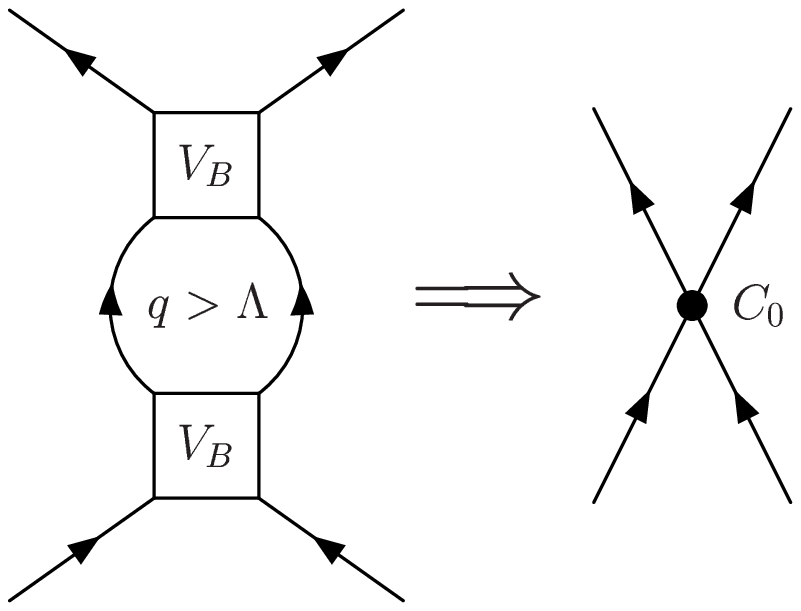}
 \caption{The equation for the $T$-matrix
   with cut-off potential $V_\Lambda$ and
   replacement of the effects of high $q$ 
   intermediate states with a contact interaction.}
 \label{fig:LSvlowk}
%\end{figure}
%
\vspace*{.2in}
% 
%\begin{figure}
 \centering 
 \includegraphics*[width=3.0in]{all7bw}
 \caption{The $^1S_0$ $V_\Lambda$ at $\Lambda=2.1\,\mbox{fm}^{-1}$
 (symbols)
   for many $\chi^2/\mbox{dof}\approx 1$ potentials \cite{SCHWENK03}.}
 \label{fig:vlowk}
\end{figure}

We illustrate the general principle by considering nucleon-nucleon scattering
in the center-of-mass frame (see Fig.~\ref{fig:LSvlowk}).  
The Lippmann-Schwinger equation iterates
a potential, which we can take as one of the $\chi^2/\mbox{dof}\approx
1$ potentials (as in Fig.~\ref{fig:vlowk}).  
Intermediate states with relative momenta
as high as $q = 20\,\mbox{fm}^{-1}$
are needed for convergence of the sum in the second term.  
Yet the elastic scattering data and the reliable long-distance
physics (pion exchange)
only constrain the potential for $q \leq 3\,\mbox{fm}^{-1}$ \cite{SCHWENK03}.

We can cut off the intermediate states at successively
lower $\Lambda$; with each step we
have to change the potential $V_\Lambda$ to maintain the same
phase shifts.
This determines the renormalization group (RG) equation
for $V_\Lambda$ \cite{SCHWENK03}.
We see in Fig.~\ref{fig:vlowk} that at $\Lambda=2.1\,\mbox{fm}^{-1}$, the
potentials have all collapsed to the same low-momentum potential
(``$V_{{\rm low\,}k}$'').
We emphasize that this potential still reproduces \emph{all} of
the phase shifts described by the original potentials.
The collapse actually occurs for $\Lambda$
between about 500 and 600\,MeV (or 2.5--3\,fm$^{-1}$), which
implies  model dependences at shorter distances.
Note that the collapse 
does \emph{not} mean that the short-distance physics is
unimportant in the $S$ channels; rather, only the coarse features are
relevant and a hard core is not needed.
(For much more discussion of $V_{{\rm low\,}k}$, including plots of the
potentials
and phase shifts in all relevant channels, see \cite{SCHWENK03}.)

The shift of each bare potential to $V_{{\rm low\,}k}$
is largely constant at low momenta, which means it is well represented 
just by
contact terms and a derivative expansion (i.e., a power series in
momentum) \cite{SCHWENK03}.
This observation illustrates explicitly
that the short-distance physics can be absorbed into local terms; this
is the essence of renormalization!
Further, it motivates the use of a local 
Lagrangian approach.
In an EFT, by varying the cutoff (or equivalent regularization
parameter), we shift contributions between loops and low-energy constants
(LEC's), just like the shift between the high-lying intermediate state
sum and the potential $V_\Lambda$.  
The long-range physics is treated explicitly 
(e.g., pion exchange)
and short-distance interactions are replaced by LEC's multiplying
contact terms 
(including derivatives).

\subsection{Effective Field Theory Ingredients}
\label{subsec:2b}

The low-energy data is insensitive to \emph{details} of short-distance
physics, so we can replace the latter with something simpler without
distorting the low-energy physics.  Effective field theory (EFT) is a
local Lagrangian, model-independent approach to this program.
Complete sets of operators at each order, determined by a well-defined
power counting, lead to a systematic expansion, which means
we can make error estimates.
Underlying symmetries are incorporated and we generate currents for external
probes consistent with the interactions.
The natural hierarchy of scales is a source of expansion parameters.

The EFT program is realized as described in
\cite{CROSSING}:
\begin{enumerate}
 \item \emph{Use the most general Lagrangian with low-energy degrees
 of freedom consistent
 with global and local symmetries of the underlying theory.}  For 
 few-nucleon chiral EFT (our example here), 
 this is a sum of Lagrangians for the zero, one,
 and two+ nucleon sectors of the strong interaction:
 \beq
   {\cal L}_{\rm eft} = {\cal L}_{\pi\pi}
             + {\cal L}_{\pi N} + {\cal L}_{NN}  \ .
 \eeq 
 Chiral symmetry constrains the form of the long-distance pion physics,
 which allows a systematic organization of pion terms \cite{VANKOLCK99}.
 
 \item \emph{Declare a regularization and renormalization scheme.} 
 For the few nucleon problem, 
 the most successful approach has been to introduce a
 smooth cutoff in momentum.  The success of the renormalization procedure
 is reflected in the sensitivity to the value of the cutoff.
 The change in observables with a reasonable variation of the cutoff
 gives an estimate of truncation errors for the EFT \cite{EPELBAUM}.
 
 \item \emph{Establish a well-defined power counting based on small expansion
 parameters.}  The source of expansion parameters in an EFT is usually 
 a ratio of scales in the problem.  For the strong interaction case,
 we have the momenta of nucleons and pions or the pion mass compared to
 a characteristic chiral symmetry breaking scale $\Lambda_\chi$, 
 which is roughly
 1\,GeV (600\,MeV is probably closer to the value in practice).
 The nuclear problem is complicated relative to $\pi\pi$ or $\pi N$
 by an additional small scale,
 the deuteron binding energy, which precludes a perturbative expansion
 in diagrams.
 Weinberg proposed to power count in the \emph{potential} and then solve
 the Schr\"odinger equation, which provides a nonperturbative summation
 that deals with the small scale.
 The NN potential is expanded as:
 \beq
   V_{\rm NN} = \sum_{\nu=\nu_{\rm min}}^{\infty}
              c_\nu Q^\nu \ ,
 \eeq
 with $Q$ a generic momentum or the pion mass, and (see \cite{CROSSING}
 for details) 
 \beq
   \nu =4-A+2(L-C)+\sum_i V_i (d_i+f_i/2-2) \ ,
 \eeq
 where the topology of the corresponding  Feynman diagram determines $\nu$.

\end{enumerate}

The organization of $V_{\rm NN}$
in powers of $Q$
is illustrated in this table
(solid lines are nucleons, dotted
lines are pions):

\begin{table}
\centering
\begin{tabular}{|c|c|c|c|}
 \hline
% table heading
   $Q^\nu$ & 1$\pi$ & $2\pi$ & $4N$ 
 \\ 
 \hline
% 1st line in table
   $ Q^0$  & 
   {%
      \raisebox{-3.5ex}{\includegraphics*[width=0.55in,angle=0.0]%
         {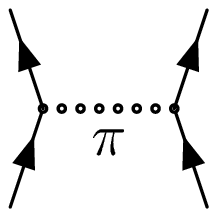}}} &
   {{---}} & 
   {%
      \raisebox{-3.5ex}{\includegraphics*[width=0.8in,angle=0.0]%
         {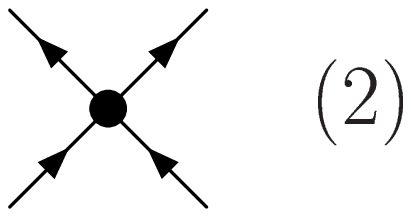}}} 
 \\[10pt]
% 2nd line in table
   {$Q^1$}  & 
   \hspace*{1.0in}  &  
   \hspace*{1.0in} & 
   \hspace*{1.0in}  
 \\[10pt]
% 3rd line in table
  {$Q^2$}  & 
  {\raisebox{-5ex}%
    {\includegraphics*[width=0.7in,angle=0.0]{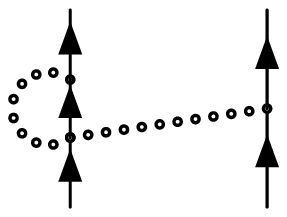}}} &   
  {\raisebox{-5ex}%
   {\includegraphics*[width=0.55in,angle=0.0]{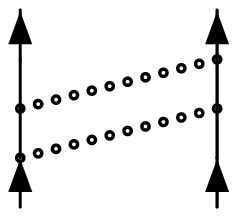}}} & 
  {\raisebox{-5ex}%
    {\includegraphics*[width=0.9in,angle=0.0]{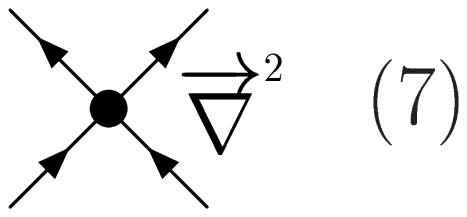}}} 
 \\[10pt]
% 4th line in table
  {$Q^3$}  &  
  {\raisebox{-4ex}%
    {\includegraphics*[width=0.75in,angle=0.0]{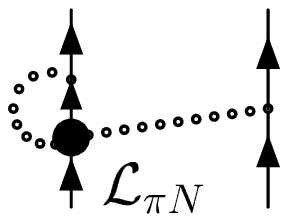}}} &   
  {\raisebox{-4ex}%
    {\includegraphics*[width=0.65in,angle=0.0]{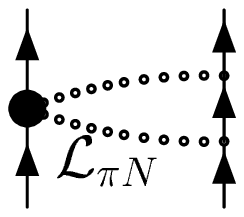}}} &  
 \\[15pt] 
 \hline
\end{tabular}
\label{tab:one}
\end{table}

\noindent
The table indicates what gets added at leading order (LO),
next-to-leading order (NLO), and next-to-next-to-leading order
(NNLO or N$^2$LO).
The low-energy constants at LO and NLO, which are the coefficients
of the $4N$ contact terms (nine total), are determined by matching
to the phase-shift data at energies up to  100\,MeV; phases at
higher energies are \emph{predictions}.  (There is also input from
$\pi N$ scattering through ${\cal L}_{\pi N}$ at NNLO.)

\vspace*{-.2in}

\begin{figure}[h]
  \centering
  \includegraphics*[width=2in,angle=0.0]{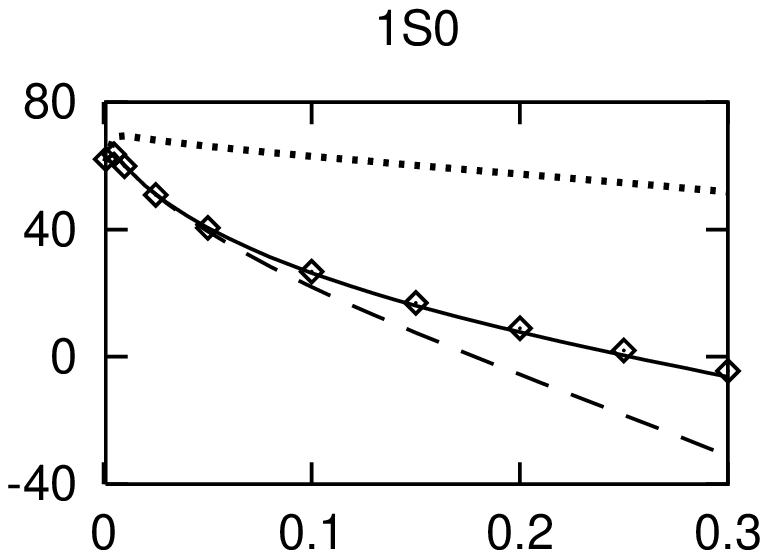} 
  \includegraphics*[width=2in,angle=0.0]{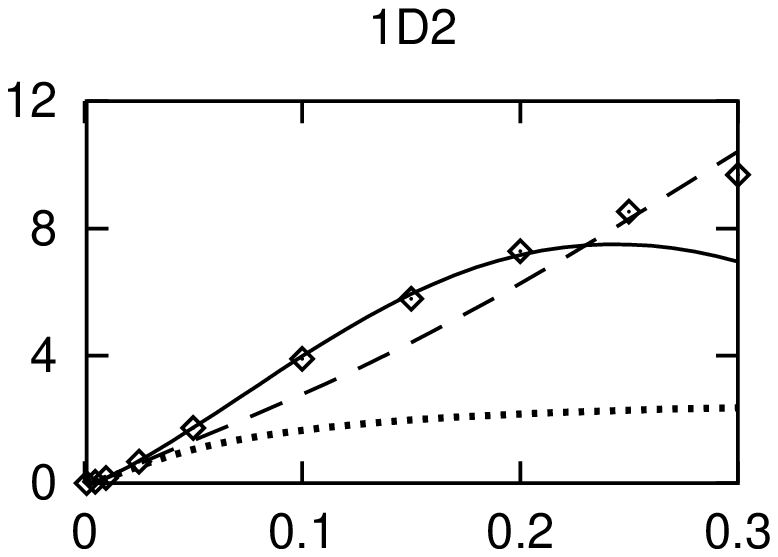}
  \caption{Representative phase shifts at LO (dots),
  NLO (dashes), and NNLO (solid) as a function of lab
  energy (in GeV) \cite{EPELBAUM}.}
  \label{fig:phases}
\end{figure}

In Fig.~\ref{fig:phases}, the systematic improvement from
LO to NNLO is evident \cite{EPELBAUM}.  
\emph{This type of systematic improvement is what we seek for the 
many-body problem.}
Recent calculations at N$^3$LO by Entem and Machleidt \cite{ENTEM03} and by
Epelbaum et al.\ \cite{EPELBAUM03}
show continued improvement.  The work by Epelbaum
et al.\  
is particularly noteworthy in providing error bands based
on varying the cutoff.  All of the phase shifts are consistently
predicted within the estimated truncation error.
Entem and Machleidt fine-tune their potential to achieve $\chi^2 \approx
1$ for the phase-shift data up to 300\,MeV lab energy.  
This puts the potential on the same footing
as conventional potentials, but sacrifices the controlled systematics of
the EFT.

An important feature of the chiral EFT is that it exhibits naturalness.
That means that when the relevant dimensional scales for any given
term are identified and factored out, 
the remaining dimensionless parameters are
of order unity.  If this were not the case, then a
systematic hierarchy would be in jeopardy.
The appropriate scheme for low-energy QCD,
which Georgi and Manohar \cite{GEORGI84b} 
called ``naive dimensional analysis'' (or NDA), assigns powers of
$f_\pi = 93\,\mbox{MeV}$ 
and $\Lambda_\chi$ to a generic term in the Lagrangian according to:
\beq
     {\cal L}_{\chi\,\rm eft}
     = c_{lmn} \left( \frac{N^\dagger(\cdots)N}{\fpi^2 \Lambda_\chi}
               \right)^l
               \left(\frac{\pi}{\fpi}\right)^m
               \left(\frac{\partial^\mu,m_\pi}{\Lambda_\chi}\right)^n
               \fpi^2\Lambda_\chi^2 \ ,
       \label{eq:NDA}
\eeq
where $l$, $m$, and $n$ are integers.

The dimensionless constants resulting when 
the LO and NLO constants are scaled this way 
\cite{EPELBAUM}
are given in the table below
for cutoffs ranging from
500 to 600\,MeV.
We see that $1/3 \lesssim c_{lmn} \lesssim 3$ in all cases, with one
exception, which implies they are natural (and the expansion is under
control).  
The one exception is $\fpi^2 C_T$, which is unnaturally
\emph{small}.  This is often a signal that there is a symmetry, and in
this case a corresponding
symmetry has indeed been identified:  the Wigner $SU(4)$
spin-isospin symmetry \cite{EPELBAUM}.

\newcommand{\fpilamsq}{f_\pi^2 \Lambda_\chi^2}

{
\renewcommand{\tabcolsep}{6pt}
\renewcommand{\arraystretch}{1.2}
\begin{table}
 \centering
 \begin{tabular}{|c|c|c|c|}
   %  \hline
   %           &  NLO              & NNLO   \\
   \hline
   $f_\pi^2 \,{C}_S$ & $-1.079 \ldots -0.953$ &
     {$f_\pi^2 \,{C}_T$} & {$0.002 \ldots 0.040$} \\
   \hline
   $\fpilamsq \,{C}_1$ & $3.143 \ldots 2.665$ &
     $4 \,\fpilamsq \,{C}_2$ & $2.029 \ldots 2.251$ \\
   $\fpilamsq \,{C}_3$ & $0.403 \ldots 0.281$ &
     $4\, \fpilamsq \,{C}_4$ & $-0.364 \ldots -0.428$ \\
   $2\, \fpilamsq \,{C}_5$ & $2.846 \ldots 3.410$ &
     $\fpilamsq  \,{C}_6$ & $-0.728 \ldots -0.668$ \\
   $4\, \fpilamsq \,{C}_7$ & $-1.929 \ldots -1.681$ & & \\ \hline
 \end{tabular}
 \label{tab:two}
\end{table}
}

We can also explore the connection between the chiral EFT and
conventional potentials by assuming that the boson-exchange potentials provide
models of the short-distance physics that is unresolved in chiral EFT
(except for the pion).
Thus this physics should be 
encoded in coefficients of the contact terms.  We can
reveal the physics by expanding the boson propagators:

\centerline{\includegraphics*[width=3.5in,angle=0.0]{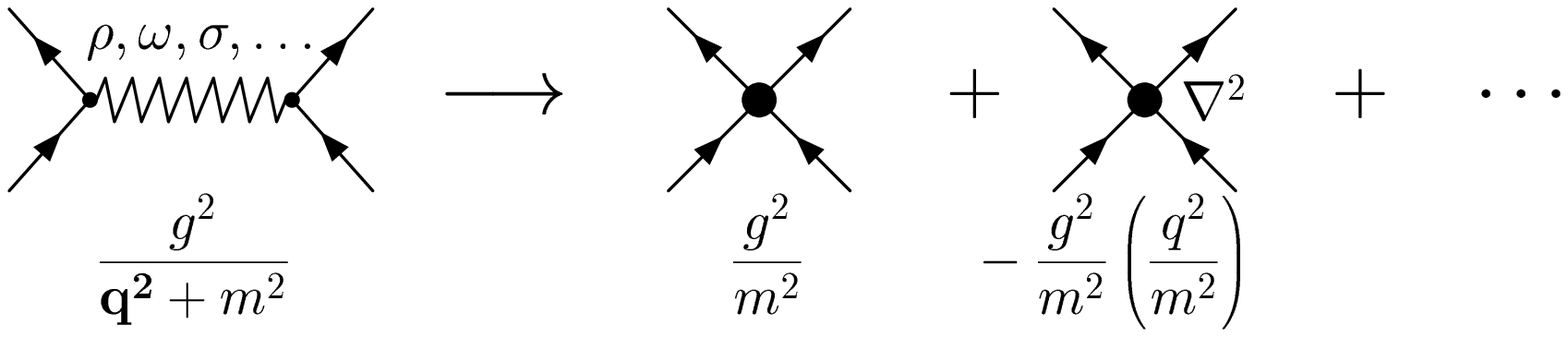}}

\noindent
By comparing to (\ref{eq:NDA}) with $m \approx \Lambda_\chi$, we
see that $g \sim \Lambda_\chi/\fpi$, which explains the large couplings
for phenomenological one-boson exchange! 

\begin{figure}
    \centering
    \includegraphics*[width=3.8in,angle=0.0]{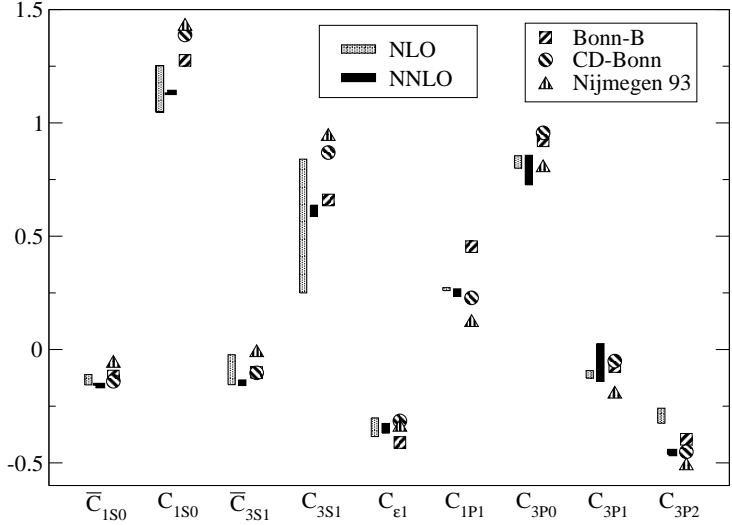}
    \caption{Comparison of EFT and one-boson-exchange coefficients
    \cite{EPELBAUM}.}
    \label{fig:res_satd}
\end{figure}

\noindent
In Fig.~\ref{fig:res_satd}, coefficients from the chiral EFT
are compared to those obtained from such an expansion applied to several
boson-exchange potentials \cite{EPELBAUM}.
The semi-quantitative agreement is remarkable and shows that there
are strong similarities between the chiral potential and standard
phenomenology.
It is not yet clear, however, whether one can conclude that 
the potentials contain reasonable models of
short-distance physics;
the agreement may just reflect the fact that they all fit the same data.

In summary, the chiral EFT for two-nucleon physics (and few-body nuclei)
is maturing rapidly.  The systematic improvement and model independent
nature is compelling.
We seek the same characteristics for our description of heavier nuclei
using covariant energy functionals.

%%%%%%%%%%%%%%%%%%%%%%%%%%%%%%%%%%%%%%%%%%%%%%%%%%%%%%%%%%%%%%%%%%%%%
%%%%%%%%%%%%%%%%%%%%%%%%%%%%%%%%%%%%%%%%%%%%%%%%%%%%%%%%%%%%%%%%%%%%%

\section{Relativistic versus Nonrelativistic EFT for Nuclei}

Why should one consider a \emph{relativistic} effective field theory for
nuclei?  Let us first play devil's advocate and argue the contrary
case.  The relevant degrees of
freedom for low-energy QCD  are pions and
nucleons (at very low energy even pions are unresolved).  
Nuclei are clearly nonrelativistic, since the Fermi momentum
$\kf$ is small compared to the nucleon mass.  The nonrelativistic NN EFT
described above is close to the successful nonrelativistic
potential and shows no (obvious) signs of problems.

In the past, a common argument against covariant treatments was that
they intrinsically relied on ``Z graphs,'' which implied
${\rm \overline NN}$ contributions that were far off shell.  The claim
was that those should really
come with form factors that would strongly
suppress their contribution \cite{BRODSKY84}.  
This argument does not hold water in light
of the effective theory principle we have highlighted:  the
high-energy, off-shell 
intermediate states may be incorrect, but that is fine,
since the physics can be corrected by local counterterms.  (Note: this
only works if we include the appropriate local operators!)
The modern
EFT argument against Z graphs is different: since they are 
short-distance degrees of
freedom, they should be integrated
out.  Indeed, it is often said
that the heavy meson fields ($\sigma$, $\omega$, and so on)
should be integrated out as well.  

This argument is sometimes advanced as a matter of principle, but the
underlying reason is the need for a well-defined power counting.
The early experience with chiral perturbation theory using relativistic
nucleons was that chiral power counting is spoiled by unwanted factors of
the nucleon mass that come essentially from Z graphs \cite{GASSER88}.  
Integrating out heavy degrees of freedom moves these large
scales to the denominators and then all is well \cite{JENKINS91}.

We'll resolve the issue of power counting in the next subsections.
But first a brief (in the legal
sense) on the side of the (covariant)
angels.  Arguments in favor of covariant approaches to nuclear structure
are collected in \cite{FURNSTAHL00} and \cite{FURNSTAHL00a}; this is
merely a skeletal summary.
For nuclear structure applications, the relevance of relativity is
\emph{not} the need for relativistic kinematics but that a covariant
formulation maintains the distinction between scalars and vectors (more
precisely, the zeroth component of Lorentz four-vectors).
There is compelling evidence that representations with large scalar and
vector fields in nuclei, of order a few hundred MeV, provide simpler and
more efficient descriptions than nonrelativistic approaches that hide
these scales.  The dominant evidence is the spin-orbit splittings.
Other evidence includes the density dependence of optical potentials,
the observation of approximate pseudospin symmetry, correlated two-pion
exchange strength, QCD sum rules, and more.

\subsection{Historical Perspective: Relativistic Hartree Approximation}

\begin{figure}
   \centering
   \includegraphics*[width=3.0in,angle=0]{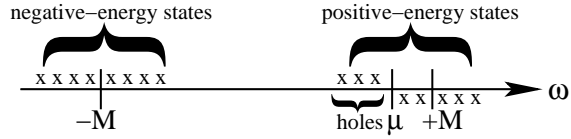}
   \caption{Pole structure for the single-particle propagator.}
   \label{fig:fig_poles}
\end{figure}

From the first applications of relativistic field theory to nuclear
structure (``quantum hadrodynamics'' or QHD), the Dirac sea had to be
considered.
Indeed, it was often hailed as being new physics missing from the
nonrelativistic description.  The simplest (and seemingly unavoidable)
consequence of the Dirac sea was an energy shift from filled
negative-energy states in the presence of a scalar field $\phi$
that shifts the effective nucleon mass to $\Mstar$
(i.e., a form of Casimir energy, see Fig.~\ref{fig:fig_poles}
for a schematic of shifted nucleon poles):
\beq 
  \delta H = -\sum_{{\bf k}\lambda} \bigl[
    ({\bf k}^2 + \Mstar{}^2)^{1/2} - ({\bf k}^2 + M^2)^{1/2}  
    \bigr]\,;  \qquad \Mstar \equiv M - g_s\phi
\eeq     

\noindent
This sum is divergent.  Two paths were taken to deal with the divergence.  
The ``no-sea
approximation'' simply discards the negative-energy contributions 
(and therefore $\delta H$) with
a casual argument about effective theories (we fix up
this argument in EFT language below).  

The other approach was to insist on
renormalizability of the QHD theory.  The usual prescription 
eliminated powers of $\phi$ up to $\phi^4$ in $\delta H$, 
leaving 
%by necessity 
a  finite shift in
the energy in the ``relativistic Hartree approximation'' (RHA) of
\cite{SEROT86}
\beqa
 \Delta {\cal E}_{\rm RHA} &=&  -{1\over 4\pi^2}\bigl[
  \Mstar{}^4 \ln({\Mstar\over M})
   - \mbox{first four powers of $\phi$}\bigr] \\
   \null &=&
   {5!\,\kappa_5 \over 20\pi^2} M^4 \Bigl({\phi\over\fpi}\Bigr)^5
   + {6!\,\kappa_6 \over 120\pi^2} M^4 \Bigl({\phi\over\fpi}\Bigr)^6
   + \cdots \ .
   \label{eq:seven}
\eeqa
In
accordance with Georgi-Manohar NDA, we have  
in the second line
scaled $\phi$ with a factor of $\fpi$  and introduced combinatoric
factors
and dimensionless couplings $\kappa_5$ and $\kappa_6$
of order unity (which absorb factors of $g_s \fpi/M \approx 1$).
The overall factor of
$M^4$, however, 
is much larger than the corresponding factor of $\fpi^2
\Lambda_\chi^2$ in NDA.  This is a signature that the power counting
will not be correct.

\begin{figure}
  \centering
  \includegraphics*[width=3.in,angle=0]{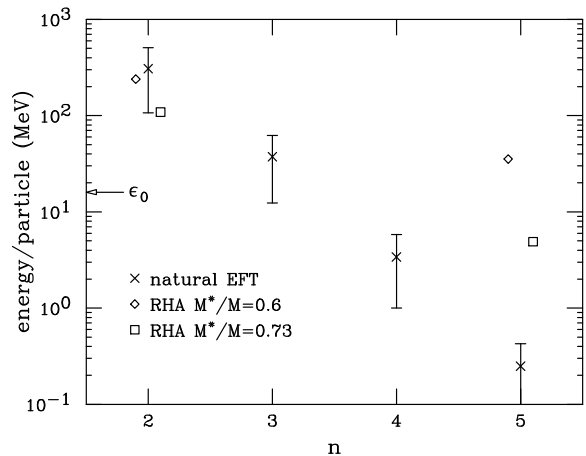}      
  \caption{Contributions to the energy per particle from terms
  in the RHA energy functional proportional to $\phi^n$
  \cite{FURNSTAHL97b}.  Two values of $\Mstar/M$ are
  considered (diamonds and squares). The 
  X's and error bars
  reflect natural NDA estimates.  }
  \label{fig:rha_compare3}
\end{figure}

In practice, 
the $\phi^5$ and higher terms are large and 
drive the self-consistent effective mass $\Mstar$ too close to $M$, 
which results in too small spin-orbit splittings (the
\textit{raison d'etre} of the relativistic approach!).
Furthermore, the loop expansion is a phenomenological disaster (two-loop
corrections qualitatively change the physics and there is no sign of
convergence \cite{FURNSTAHL89}).
The $M^4$ factor in (\ref{eq:seven}) also badly
violates $N_c$ counting that we would expect in a
low-energy theory of QCD. 
($N_c$ is the number of colors.) 
With $\Lambda_\chi$ associated with a meson
mass, $M^4 \propto (N_c)^4$ while $\Lambda_\chi^2\fpi^2 \propto
(1)^2(N_c^{1/2})^2 \propto (N_c)^1$ \cite{FURNSTAHL97b}.

In general, the RHA energy functional exhibits \emph{unnaturally} large
contributions to $\phi^n$ terms, as seen in Fig.~\ref{fig:rha_compare3}.
A natural effective theory would have decreasing contributions, so that
$\phi^5$ and higher terms would have only minor effects (easily absorbed
elsewhere).  With a value of $\Mstar/M \approx 0.6$, which is required
for realistic spin-orbit splittings, the energy/particle from the
$\phi^5$ term is two orders of magnitude too large.  The system adjusts
$\Mstar$ upward
to reduce this contribution, but the consequence is small spin-orbit
splittings.

\subsection{Power Counting Lost / Power Counting Regained}

Gasser, Sainio, and Svarc first adapted chiral perturbation theory
(ChPT) to pion-nucleon physics using relativistic nucleons
\cite{GASSER88}.
In contrast to the pion-only sector, however, the loop and momentum
expansions did not agree and systematic power counting was lost.
The heavy-baryon formulation was introduced to restore power counting
through an expansion in $1/M$, which means a nonrelativistic
formulation \cite{JENKINS91}.

In 1996, Hua-Bin Tang wrote a seminal paper 
\cite{TANG96} in which he said ``\ldots
EFT's permit useful low-energy expansions only if we absorb \emph{all}
of the hard-momentum effects into the parameters of the Lagrangian.''
The key observation he made was that: ``When we include the nucleons
relativistically, the anti-nucleon contributions are also hard-momentum
effects.''  Once the ``hard'' part of a diagram is absorbed into
parameters, the remaining ``soft'' part satisfies chiral power counting.

There are several prescriptions on the market for carrying out this
program.  Tang's original prescription (developed with Ellis) involved
the expansion and resummation of propagators \cite{TANG96,ELLIS98}.  
This basic idea was
systematized for $\pi N$ by Becher and Leutwyler \cite{BECHER99}
under the name
``infrared regularization.''  More recently, Fuchs et al.\ described an
alternative prescription using additional finite subtractions (beyond
minimal subtraction) in dimensional regularization \cite{FUCHS03}.
Goity et al.\ \cite{GOITY01} and Lehmann and Pr\'ezeau
\cite{Lehmann:2001xm}
extended infrared regularization
to multiple heavy particles.  The result for particle-particle loops
reduces to the usual nonrelativistic result for small momenta
while particle-hole loops in free space vanish identically.

\subsection{Effective Action and the No-Sea Approximation}

At the one-loop level, we will be able to apply the necessary
subtractions without specifying the regularization and renormalization
prescription in detail.  
It is convenient to use an effective action formalism to carry out the
EFT at finite density.
The effective action is obtained by functional Legendre transformation
of a path integral generating functional with respect to source terms.
This is analogous to Legendre transformations in thermodynamics.
We assume some familiarity with effective actions; see
\cite{ILIOPOULOS,NEGELE88,PESKIN95,WEINBERG96} to learn more.

QHD models with heavy-meson fields correspond to effective actions with
\emph{auxiliary fields}, introduced to reduce the fermion integration
to just a gaussian integral.  We denote the effective action 
$\Gamma[\phi, V^\mu]$, with classical
fields $\phi$ and $V^\mu$ (suppressing all other fields)
\cite{FURNSTAHL89}.
An alternative is to work with a point coupling model, in which case the
effective action will be a functional of the scalar nucleon density and
the vector current.

There will \emph{always} be a gaussian fermion integral, which yields a
determinant that appears in the effective action as the trace over
space-time and internal variables of the logarithm of a differential
operator:
\beq
 \Tr\!\ln(i\!\!\not\!\partial
               + \mu\gamma^0
               - \Mstar - \gv\!\!\not\! V) \equiv \Tr\!\ln G^{-1}
      \ .
      \label{eq:eight}
\eeq
In the point-coupling case, optical potentials 
that are functions of the density and current
appear directly rather
than through meson fields, but the form is the same as (\ref{eq:eight}).
This $\Tr\!\ln$ depends on $\mu$, the nucleon chemical potential.  

The derivative expansion of the $\Tr\!\ln$ at $\mu=0$ takes the 
form \cite{PERRY86}:
\beqa
 -i\, \Tr\!\ln(i\!\!\not\!\partial
               - \Mstar - \gv\!\!\not\! V) &=&
      \int\! d^4x\ \bigl[ U_{\rm eff}(\phi) 
        + \frac{1}{2}Z_{1s}(\phi)\partial_\mu\phi\partial^\mu\phi
      \nonumber\\
       & & \quad\null  + \frac{1}{2}Z_{2s}(\phi)(\square\phi)^2
         + \cdots \bigr] \ ,
\eeqa
which shows that this is a \emph{purely local} potential in the meson fields.
This means it can be entirely absorbed into local terms in the
Lagrangian.  That is, by adjusting the constants appropriately, this
contribution to the energy will not appear.
We specify a consistent subtraction at
a specific $\mu$, which means removing a set of local terms (that are
implicitly absorbed by parameter redefinition).  The obvious choice is
$\mu = 0$.

We can accomplish this in practice at $\mu\neq 0$
simply by performing a  subtraction
  of the $\Tr\!\ln$ evaluated at $\mu=0$.  
Note that this is \emph{not} a vacuum
subtraction, because it depends on the finite-density background fields
$\phi(x)$ and $V^\mu(x)$.  It is simply a choice for shifting ``hard''
Dirac sea physics into the coefficients.
We emphasize that the \emph{same} coefficients in the derivative
expansion are subtracted for any background fields.
For the ground state, this will simply remove all explicit evidence of the 
negative-energy states.
But there are fixed consequences for treating linear response (RPA)
that still involve negative-energy states
\cite{FURNSTAHL87,SHEPARD89,DAWSON90,Furnstahl:2002fz}.

For the ground state, field equations for the meson potentials are obtained
by extremizing $\Gamma[\phi,V^\mu]$.  These equations determine static
potentials $\phi_0({\bf x})$ and $V_0({\bf x})$ for the ground state;
the corresponding Green's function is the Hartree propagator, denoted
$G_H$.
Then $\Gamma[\phi_0({\bf x}),V_0({\bf x})]$ is proportional to the
zero-temperature thermodynamic potential $\Omega = E - \mu N$.
Both $+ i\,\Tr\!\ln \GHinv(\mu)$ and  $+ i\,\Tr\!\ln \GHinv(0)$
are diagonal in the \emph{same} single-particle basis 
$\psi_\alpha({\bf x}) e^{i\omega x_0}$, where the $\psi_\alpha({\bf x})$'s
are solutions  with eigenvalues $\epsilon_\alpha$
to the Dirac equation in the static potentials.
Thus the subtraction is simple ($T$ is a constant time that cancels out):
\beqa
  \Gamma/T &=& -\Omega = \left[{- i\,\Tr\!\ln \GHinv(\mu)}
           + {i\,\Tr\!\ln \GHinv(0)} + \cdots\right]/T
      \nonumber\\
   &=& \sum_\alpha^{\epsilon_\alpha<\mu} (\mu -\epsilon_\alpha)}
           - {\sum_\alpha^{\epsilon_\alpha<0} (-\epsilon_\alpha)}
              + \cdots
%           - \mbox{[vac.\ sub.]} 
           = {\sum_\alpha^{0<\epsilon_\alpha<\mu} (\mu -\epsilon_\alpha)
           \ ,
\eeqa
where we have omitted the parts of $\Gamma$ (and $\Omega$) 
depending only on meson potentials, and the vacuum subtraction 
for the baryon number \cite{FURNSTAHL95,Furnstahl:2002fz}.
Thus the contribution to the energy is simply a sum over occupied,
positive-energy, single-particle eigenvalues.
This is the ``no-sea'' prescription! 

Similarly, we can calculate the scalar density (the expectation value
of $\overline \psi\psi$) from the effective
action, which yields
\beq
  \rho_s({\bf x}) \propto 
     \delta\Gamma/\delta(g_s\phi({\bf x})) \Longrightarrow
      \Tr G_H(\mu)-\Tr G_H(0)  \ .  
\eeq
The subtraction removes the contribution from the sum over
negative-energy states.
This is indicated diagrammatically as :

\vspace*{-.1in}

\begin{figure}[h]
  \centering
  \includegraphics*[width=4.0in,angle=0]{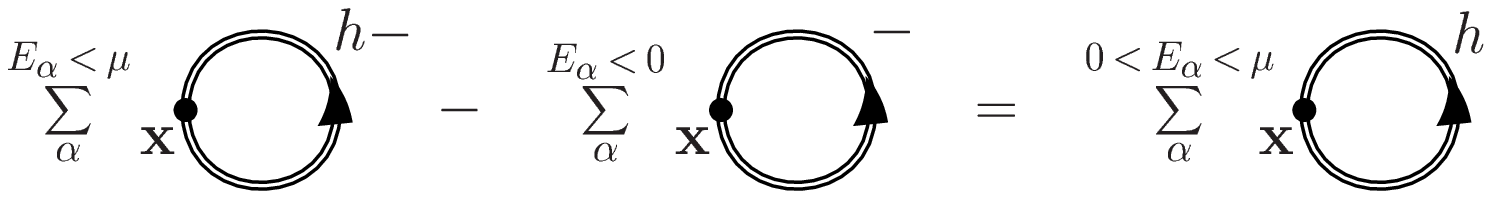}
  \label{fig:fig_rhos2}
\end{figure}

\vspace*{-.1in}

\noindent
where ``$h$'' denotes holes and
``$-$'' denotes negative-energy states.
Thus, we recover the ``no-sea approximation'' for the ground state.
What is the corresponding result for RPA excitations?

Consider $\Gamma[\phi,V^\mu]$ evaluated 
with \emph{time-dependent} fluctuations
about the static ground-state potentials:
\beq
   \phi(x) = \phi_0({\bf x}) + {\wt\phi(x)}  \ ,
   \qquad
   V^\mu(x) = V_0({\bf x})\,\delta_{\mu 0} + {\wt V^\mu(x)} \ .
\eeq
We can access the linear response directly from the effective action, 
because $\Gamma$ is the
generator of one-particle-irreducible Green's functions \cite{PESKIN95}.
The linear response is dictated by the second-order terms
in powers of $\wt\phi,\wt V^\mu$.  So we expand
\beq
   \Gamma =  - i\,\Tr\!\ln G^{-1}(\mu)
           + i\,\Tr\!\ln G^{-1}(0) + \ldots
           \label{eq:thirteen}
\eeq
in $\wt\phi,\wt V^\mu$ (this generates the ``polarization insertion''):
\beq
  \ln(G_H^{-1} + g_s\wt\phi)
           = \ln(G_H^{-1}) \bigl[\ldots  
               - \frac{1}{2} G_H\, (g_s\wt\phi)\, G_H\,
              (g_s\wt\phi)
                     + \ldots\bigr]  \ .
\eeq
The two $G_H$'s mean that we generate rings with Hartree propagators:
\beq
  - i\,\Tr\!\ln G^{-1}(\mu) \Longrightarrow
  \raisebox{-0.2ex}{\parbox{1.in}{%
     \includegraphics*[width=1.in,angle=0]{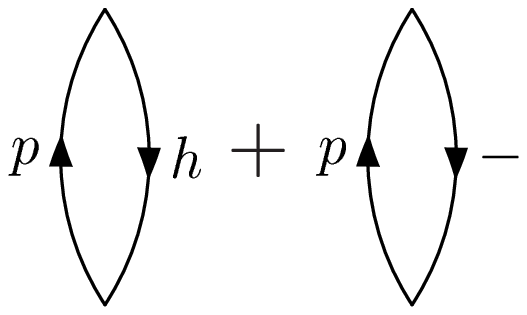}}}
    \quad
   - i\,\Tr\!\ln G^{-1}(0) \Longrightarrow
  \raisebox{-0.2ex}{\parbox{0.45in}{%
  \includegraphics*[width=0.45in,angle=0]{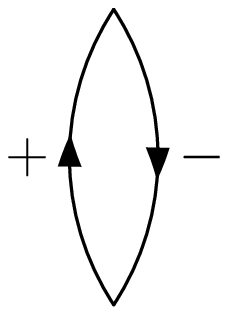}}}
\eeq 
but different spectral content.  Combining these contributions as in
(\ref{eq:thirteen}):
\begin{figure}[h]
  \includegraphics*[width=4in,angle=0]{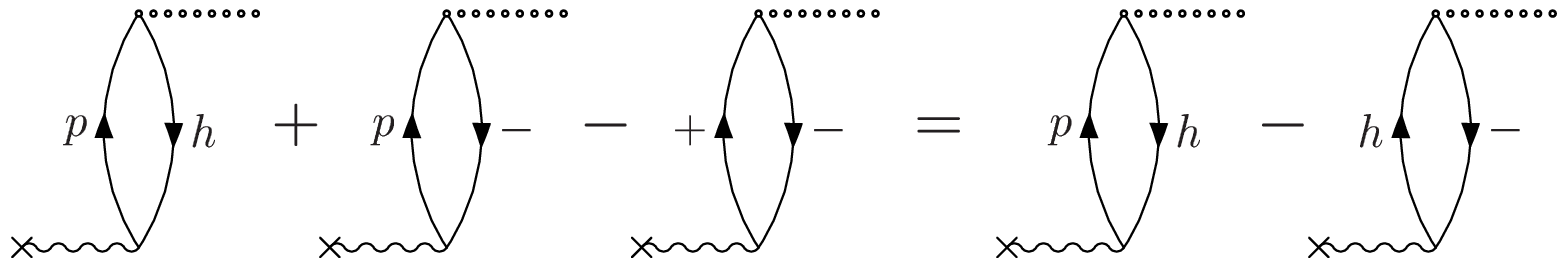}
  \centering
\end{figure}

\noindent
We see that the consistent prescription for the RPA is to include
particle-hole pairs ($ph$) \emph{and} pairs formed from holes and
negative-energy states ($h-$).

The lesson from effective field theory is that we must absorb
hard-momentum Dirac-sea physics into the parameters of the energy
functional.  
This explains the successful relativistic mean-field and RPA ``no-sea''
phenomenology.
The QCD vacuum effects are automatically encoded in the fit parameters;
there is no need to refit at different densities.
In principle, to absorb all of the short-distance physics
we need \emph{all} possible counterterms.
In practice, we rely on naturalness to make truncations (which justifies
the success of conventional models with limited terms).

The effective action formalism is a natural framework for systematically
extending the one-loop treatment.
By embedding the ``mean-field'' models 
within the density functional theory (DFT) framework,
we realize that
correlations are approximately included with the present-day mean-field
functionals, which  can be generalized to systematically improve them
according to EFT power counting.   Note that the effective action
formulation leads naturally to time-dependent DFT as well.
To carry out this program, a generalized infrared regularization scheme 
(or equivalent, possibly with a cutoff) is needed.

%%%%%%%%%%%%%%%%%%%%%%%%%%%%%%%%%%%%%%%%%%%%%%%%%%%%%%%%%%%%%%%%%%%%%
%%%%%%%%%%%%%%%%%%%%%%%%%%%%%%%%%%%%%%%%%%%%%%%%%%%%%%%%%%%%%%%%%%%%%

\section{Toward More Systematic Energy Functionals}

The energy functionals of relativistic mean-field models need
improvement in a number of areas.  As stressed in our first discussion
of EFT's, long-range physics must be included explicitly.
However, chiral physics as manifested in
long-range pion interactions is not included explicitly 
in mean-field models and the treatment of
long-range correlations is unlikely to be adequate.  
Pairing is still more
an art than a science.  Isovector physics in general has problems, as
asymmetric ($N \neq Z$) nuclear matter is poorly constrained. 
Here we focus on the last of these. 

\begin{figure}
  \centering
  \includegraphics*[width=3.in]{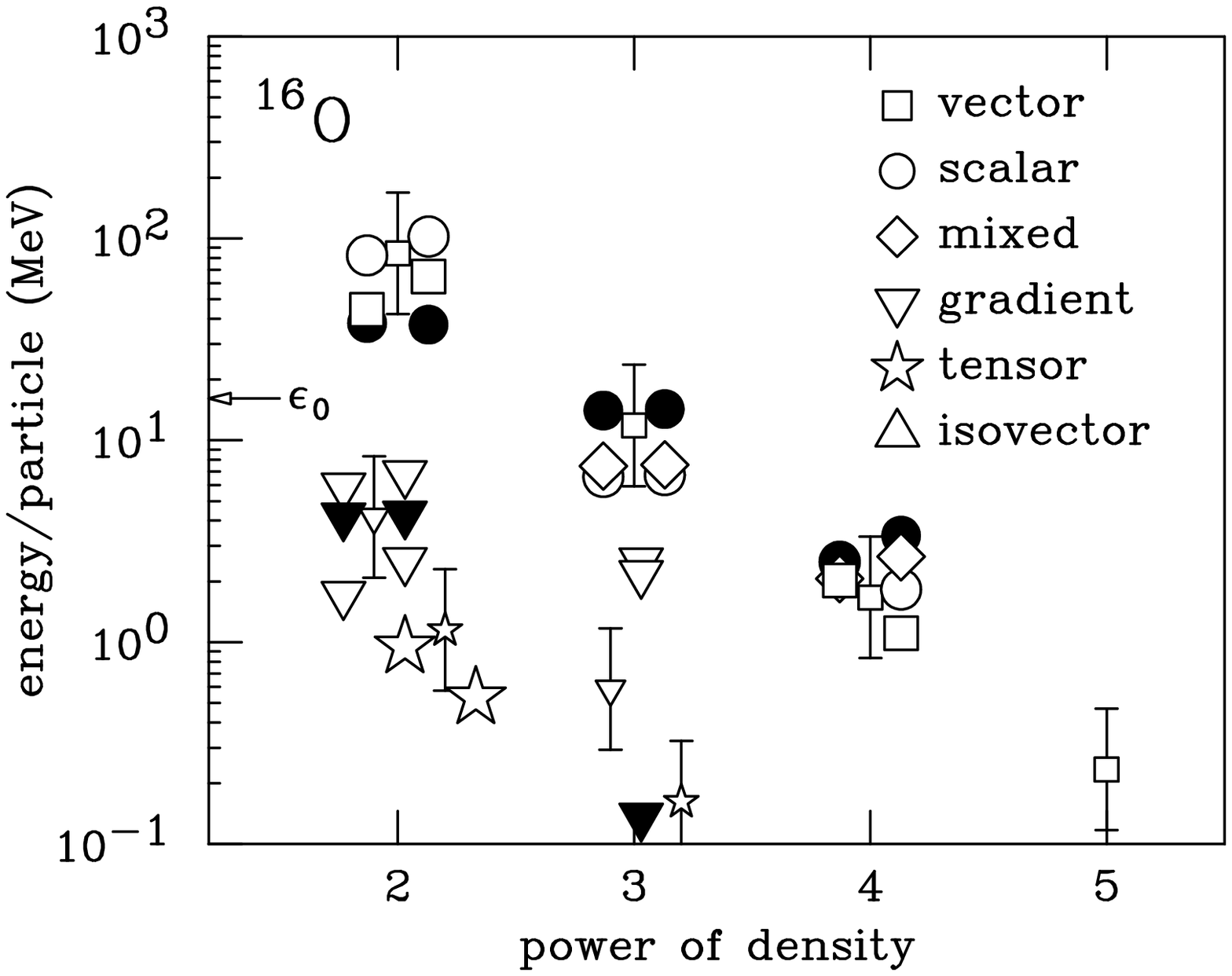}
  \includegraphics*[width=3in]{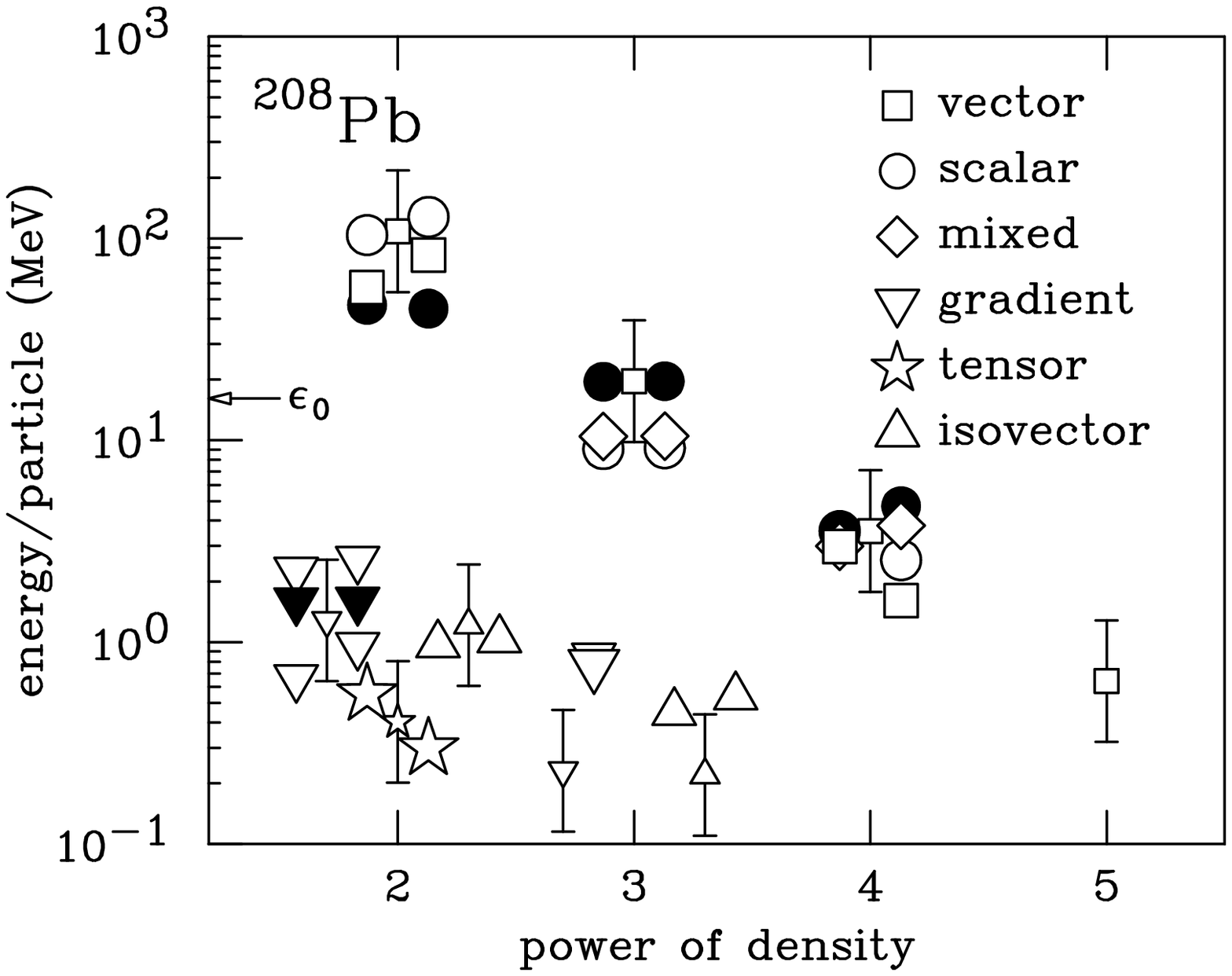}
  \caption{Estimates in finite nuclei using averages of densities and
  results from two models (at left and right of error bars)
  \cite{FURNSTAHL00a}.}
  \label{fig:be_pc_o16}
\end{figure}

EFT-inspired power counting (in the form of NDA \cite{FRIAR96}) 
can be applied
to mean-field energy functionals to quantify the isovector
constraints.
Estimates of individual contributions in a given functional to
the energy per particle of a finite nucleus
can be made using NDA rules to associate powers of $\fpi$
and $\Lambda_\chi$ with scalar ($\rho_s$), baryon ($\rhoB$),
isovector ($\rho_3$), and tensor ($\tensor$) densities and gradients:
\beq
  \rhost,\rhoBt\rightarrow 
	     {\ds\langle\rhoB\rangle \over \ds\ts\fpi^2\Lambda}
  \quad 
  \wt\grad\rhoB\rightarrow 
	      {\ds\langle\grad\rhoB\rangle \over \ds\ts\fpi^2\Lambda^2}
  \quad
  \wt\tensor\rightarrow
	    {\ds\langle\tensort\rangle\over\ds\ts\fpi^2\Lambda}
  \quad
  \rhothreet\rightarrow {\ds Z-N\over \ds\ts 2A}
	       {\ds\langle\rhoB\rangle\over\ds\ts\fpi^2\Lambda}
\eeq
and then using
local density approximations to estimate the densities.
Typical results are shown in Fig.~\ref{fig:be_pc_o16}.
The hierarchy of isoscalar contributions is clear and shows that
three orders of parameters can be constrained
(contributions below about 1\,MeV are not constrained).
In contrast,
the isovector parameters are weakly constrained by the energy;
essentially only one parameter is determined.
Naturalness implies that the $(\grad\rhothree)^2$ contribution to
observables such as the skin thickness $r_n-r_p$ is small
(see \cite{FURNSTAHL02b}), so we can
focus on terms in the functional that contribute for uniform systems.
This leads us to examine the symmetry energy in uniform matter.

\subsection{Power Counting and the Symmetry Energy}

The relationship between the neutron skin in $^{208}$Pb and the symmetry
energy in uniform matter has been studied in detail 
\cite{OYAMATSU98,BROWN00,TYPEL01,FURNSTAHL02b,DANIELEWICZ03}.
We can frame the discussion in terms of the energy per particle 
$E(\rho,\alpha)$ in asymmetric matter, which can be expanded as:
\beq
 E(\rho,{\alpha}) 
   = E(\rho,{0}) 
       + S_2(\rho) {\alpha^2} + S_4(\rho) {\alpha^4}
       + \cdots   \qquad {\alpha \equiv {N-Z \over A}}
       \label{eq:seventeen}
\eeq
where $\rho$ is the baryon density and $\alpha$ is an asymmetry
parameter.
The $\alpha=0$ term has the usual expansion around the saturation
density $\rho_0$:
\beq
 E(\rho,0) = - { a_{\rm v}} + {K_0 \over 18 \rho_0^2} (\rho - \rho_0)^2 
    + \cdots
\eeq
while the quadratic symmetry energy $S_2(\rho)$ is expanded as:
\beq
 {S_2(\rho)} = {a_4} + {{p_0}\over \rho_0^2} (\rho - \rho_0)
           + {\Delta K_0 \over 18 \rho_0^2} (\rho - \rho_0)^2 + \cdots
\eeq
Two key parameters are the symmetry energy at saturation, $a_4$, and the
linear density dependence, $p_0$.
(Note: $S_4(\rho)$ is unimportant numerically.)

\begin{figure}
  \centering
  \includegraphics*[angle=-90.0,width=3.1in]{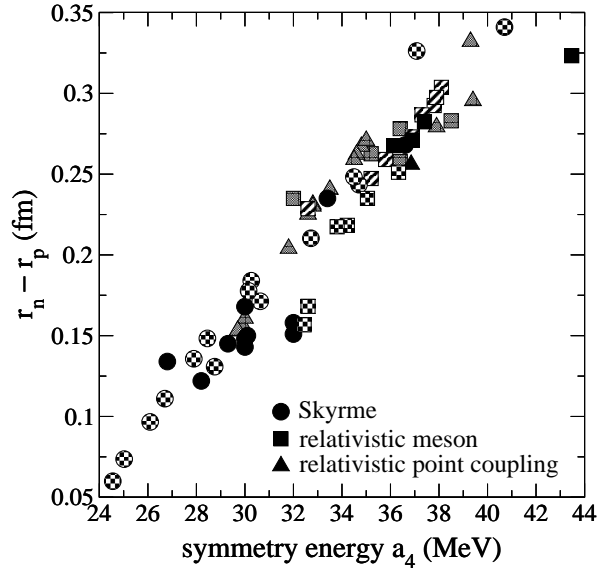}
  \caption{Skin thickness in $^{208}$Pb versus 
  the equilibrium symmetry energy \cite{FURNSTAHL02b}.}
  \label{fig:rnrp_vs_a4}
\end{figure}

\begin{figure}[h]
  \centering
  \includegraphics*[angle=0.0,width=3.1in]{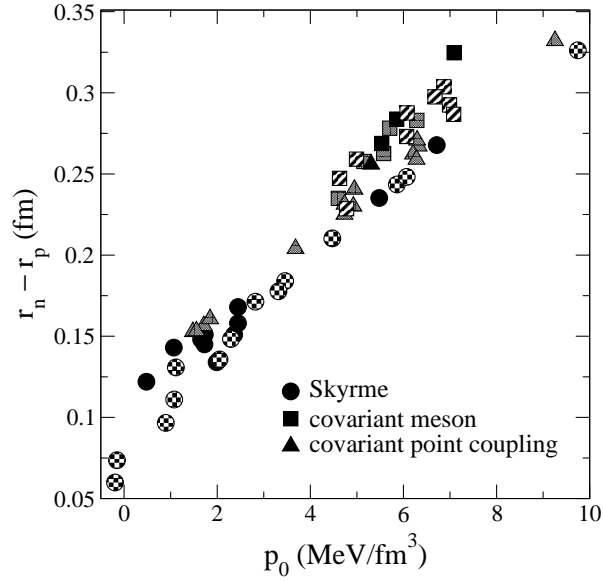}
  \caption{Skin thickness in $^{208}$Pb as a function
    of the density dependence of the symmetry energy \cite{FURNSTAHL02b}.}
  \label{fig:rnrp_vs_p0}
\end{figure}

A study of a wide range of mean-field models, both nonrelativistic
Skyrme and relativistic mean-field (meson-based and point-coupling),
show strong correlations between the skin thickness in $^{208}$Pb
and the symmetry energy parameters.
In Figs.~\ref{fig:rnrp_vs_a4} and \ref{fig:rnrp_vs_p0}, this correlation
is shown explicitly.
While there is a large spread in the values of $S_2(\rho_0) = a_4$,
if we plot the spread (i.e., the standard deviation) as a function
of $\rho$, we see that the spread
is smallest 
at a density corresponding to the \emph{average} density in heavy
nuclei (see Fig.~\ref{fig:S2sd_vs_rho}).  
This result is consistent with our conclusion from power counting that
only one parameter is determined: it is the symmetry energy at some lower
density.

\begin{figure}
  \centering
  \includegraphics*[angle=0.0,width=2.2in]{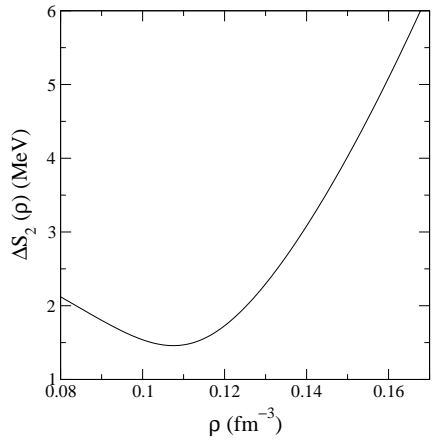} 
  \caption{The spread (standard deviation) of $S_2(\rho)$ [see
  (\ref{eq:seventeen})] for many mean-field models fit to nuclear
  bulk properties \cite{FURNSTAHL02b}.}
  \label{fig:S2sd_vs_rho}
\end{figure}

Since $p_0$ is a pressure that pushes neutrons out
against the surface tension in $^{208}$Pb, it is instrumental
in determining the skin thickness \cite{Horowitz:2001ya}.
How well do we know its value?  
A comparison of symmetry energies predicted
by realistic non-relativistic potentials was made
in \cite{MACHLEIDT}
with the result in Fig.~\ref{fig:machleidt}.
Reading off the slopes implies that $p_0 \approx 2.1\,\mbox{MeV/fm}^{-3}$
with only a 10\% spread.
From relativistic RBHF calculations, $p_0 \approx
3.3\,\mbox{MeV/fm}^{-3}$.
In either case, there is a strong discrepancy with conventional meson
mean-field models, which have $p_0\geq 4$. 

\begin{figure}
  \centering
  \includegraphics*[angle=0,width=3.in]{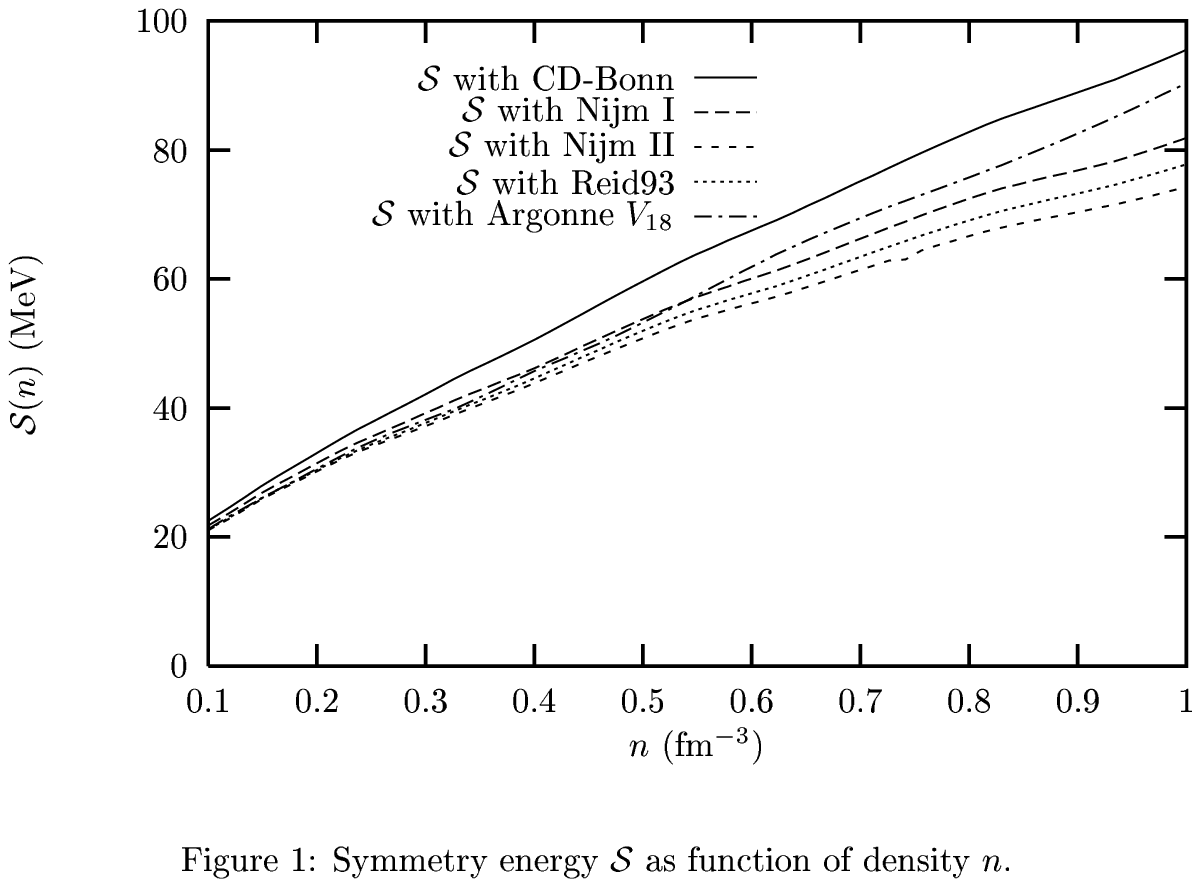}
  \caption{Comparison of the symmetry energy for realistic potentials
  \cite{MACHLEIDT}.}
  \label{fig:machleidt}
\end{figure}

This conclusion is supported by
Danielewicz \cite{DANIELEWICZ03},
who has analyzed the relationship in terms of
surface vs.\ volume symmetry energies.  Using mass formula constraints
in conjunction with skin thickness analysis, he concludes
that $0.56 \leq S_2(\rho_0/2)/a_4 \leq 0.81$, which allows the models
between the vertical lines in Fig.~\ref{fig:rnrp_vs_s}.
Once again, conventional relativistic models are excluded.

\begin{figure}
  \centering
  \includegraphics*[width=3.1in,angle=0.0]{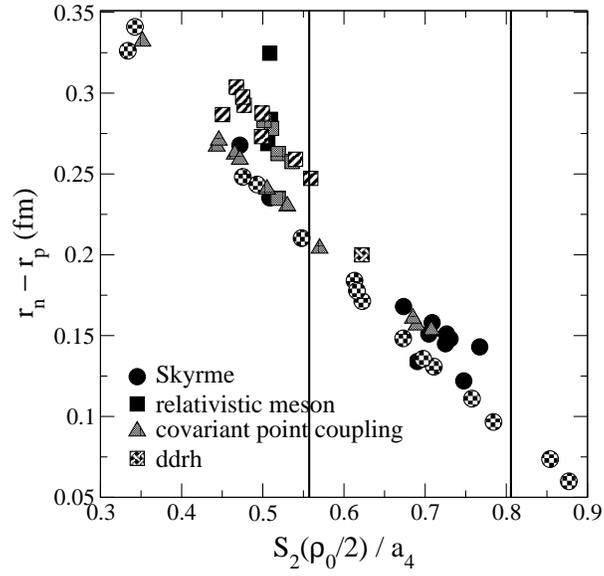}
  \caption{Skin thickness in $^{208}$Pb versus
    the normalized density dependence of the symmetry 
    energy.}
  \label{fig:rnrp_vs_s}
\end{figure}

There are several ways out.  Point coupling \cite{NIKOLAUS97,RUSNAK97}, 
DDRH \cite{Fuchs:1995as, Niksic:2002yp}, 
and new relativistic
mean-field models (with mesons) \cite{Horowitz:2001ya}
get lower values of $p_0$ by adding
isovector parameters (implicitly, in the case of DDRH).
However, they achieve lower $p_0$ from an unnatural cancellation 
between different orders according
to the present power counting \cite{FURNSTAHL02b}.  
This doesn't mean it is wrong, but it
means we haven't captured the physics of the symmetry energy.  
For example, it
might be pion physics with a different functional form.  
These results imply that we should revisit the
symmetry energy in covariant models within a chiral EFT.

\subsection{Guidance from DFT for Solid-State or Molecular Systems}

The density functional theory (DFT) perspective helps to \emph{explain}
successful ``mean-field'' phenomenology, but DFT existence theorems are
not constructive.
NDA power counting gives guidance on truncation but not on the
analytic form of functionals.
One approach toward more accurate energy functionals would be to
emulate the quantum chemists.

The Hohenberg-Kohn free-energy for an inhomogeneous electron gas
in an external potential 
takes the form ($\densx$ is the charge density here)
\beq
   \FHK[\densx] = \Fni[\densx] +  \frac{e^2}{2} \!
        \int\! d^3x\, d^3x' 
          \frac{n({\mathbf x})n({\mathbf x'})}{|{\bf x}-{\bf x'}|}
    + \Exc[\densx] \ ,
\eeq
where the direct Coulomb piece is explicit and
the Kohn-Sham non-interacting functional is evaluated using
\beq
  \Fni[\densx] = \sum_{i=1}^A \langle \psi_i | \hat t_i | \psi_i \rangle
          = \sum_{i=1}^N \epsilon_i - \int\!d^3x\, 
    n({\mathbf x}) v_{\rm eff}({\mathbf x})  \ .
\eeq 
The $\psi_i$'s are solutions to
\beq
  \Bigl( -\frac{\hbar^2}{2m}\nabla^2 + v_{\rm eff}({\mathbf x}) 
  \Bigr) \psi_i({\mathbf x}) = \epsilon_i \psi_i({\mathbf x})
\eeq
with the property that the \emph{exact} density of the full system is given by
\beq
  n({\mathbf x}) = \sum_{i=1}^N |\psi_i({\mathbf x})|^2
  \ ,
\eeq
which is the density of a noninteracting system with 
effective potential
\beq
  v_{\rm eff} = v -e\phi + v_{\rm xc}\ , \quad \mbox{where} \quad
  v_{\rm xc}({\mathbf x}) \equiv \delta E_{\rm xc}/\delta
      n({\mathbf x}) \ .
\eeq
$v$ is the external potential, and $\phi$ is the Hartree potential.

Solving for the orbitals and energies is the easy part.  The hard part
is finding a reliable approximation to the exchange-correlation
functional $\Exc$.  The approach used by quantum chemists uses the
local density approximation
\beq
  \Exc[\densx] \approx
          \int\! d^3x \,  {\cal E}_{{\rm xc}}(\densx) \ ,
\eeq
where the energy density
${\cal E}_{{\rm xc}}$ is fit to a Monte Carlo calculation of the
homogeneous electron gas (see Fig.~\ref{fig:fig_epsilonxc}).
In practice, one uses parametric formulas, such as 
\beq
  {\cal E}_{\rm xc}(n)/n =
          -0.458/r_s - 0.0666G(r_s/11.4)
\eeq 
with
\beq
  G(x) = \frac12\left\{(1+x)^3 \log(1+x^{-1}) - x^2 + \frac12x -
                  \frac13 \right\} 
\eeq
(the ones in common use are more sophisticated).
The calculational procedure, which includes correlations 
through $E_{{\rm xc}}$,
has the form of a ``naive'' mean-field approach with the
additional potential
\beq
   v_{\rm xc} = \left.\frac{d[{\cal E}_{\rm xc}(n)]}{dn}
                \right|_{n=n({\mathbf x})}  \ .
\eeq
An example of a successful atomic
LDA calculation is shown in Fig.~\ref{fig:argon}.
Significant improvements to the LDA 
are made using (generalized) gradient expansions
\cite{PERDEW96,PERDEW99}.
The DDRH approach to relativistic energy functionals for nuclei, in which
the density dependence of a nuclear matter calculation is incorporated
through density-dependent couplings \cite{Fuchs:1995as, Niksic:2002yp}, 
is in the spirit of these
LDA plus gradient expansion approaches.

\begin{figure}[t]
  \centering
  \includegraphics*[width=3.1in,angle=0.0]{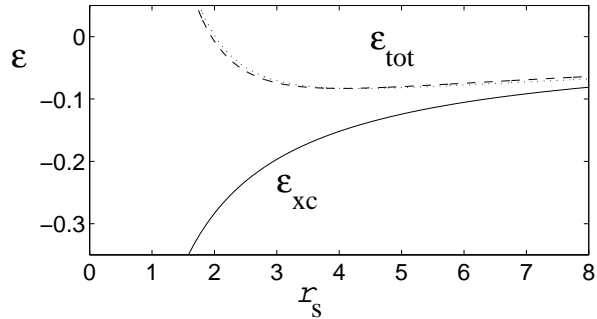}
  \caption{The exchange-correlation functional for a uniform
  system ($r_s \propto 1/\kf$) \cite{ARGAMAN00}.}
  \label{fig:fig_epsilonxc}
\end{figure}
\begin{figure}
  \centering
  \includegraphics*[width=3.1in,angle=0.0]{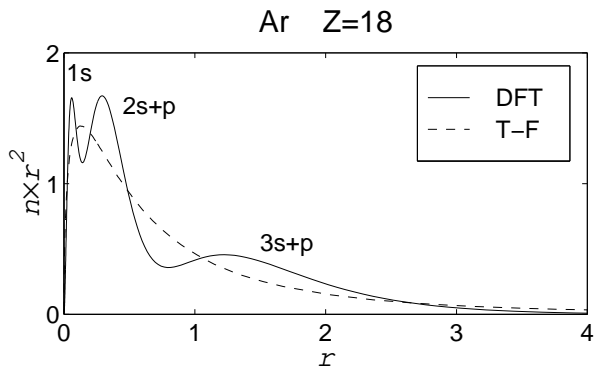}
  \caption{The charge density for argon atom calculated
  using DFT (in the LDA) and the Thomas-Fermi approximation
   \cite{ARGAMAN00}.}
  \label{fig:argon}
\end{figure}

While there are many successes to be found, it is also clear that the
situation is not entirely satisfactory.
From \emph{A Chemist's Guide to DFT} \cite{KOCH2000}: 
  \begin{quote}
    ``{To many, the success of DFT appeared somewhat miraculous,
    and maybe even unjust and unjustified.}  Unjust in view of the easy
    achievement of accuracy that was so hard to come by in the wave
    function based methods.  And unjustified it appeared to those who
    doubted the soundness of the theoretical foundations.'' 
%    {\re There has
%    been misunderstanding concerning the status of the
%    one-determinantal approach of Kohn and Sham, which superficially
%    appeared to preclude the incorporation of correlation effects.}''
  \end{quote}
From Argaman and Makov's article  \cite{ARGAMAN00} (highly recommended!):
     \begin{quote}
       ``{It is important to stress that all practical
       applications of DFT rest on essentially uncontrolled
       approximations, such as the LDA \ldots}''
     \end{quote}
and from the authors of the state-of-the-art gradient approximations
\cite{PERDEW96}:
      \begin{quote}
        ``{Some say that `there is no systematic way to construct
        density functional approximations.'}  But there are more or
        less systematic ways, and the approach taken \ldots here is
        one of the former.''
      \end{quote}
Thus we are motivated to consider a more systematic approach: EFT!

\subsection{DFT in Effective Field Theory Form}

It is particularly natural to consider DFT in thermodynamic terms, in analogy
to finding via Legendre transformation
the energy as a function of particle number from the
thermodynamic potential as a function of the chemical 
potential \cite{ARGAMAN00}.
This analogy implies the effective action
formalism will be well suited.
The EFT will provide power counting for diagrams and gradient
expansions, thereby systematically approximating a model-independent
functional.
The EFT tells us how to renormalizes divergence consistently 
and gives insight into analytic structure of the functional.

As a laboratory for exploring DFT/EFT, we can consider a dilute
``natural'' Fermi gas in a confining potential, which is realized
experimentally by cold atomic Fermi gases in optical traps.
Natural in this context means that the scattering length is comparable
to the range of the potential (rather than being fine tuned
to a large value).
The EFT Lagrangian takes the form
 \beq
  {\cal L}_{\rm eft} = 
               \psi^\dagger \bigl[i\frac{\partial}{\partial t} 
               + \frac{\nabla^{\,2}}{2M}\bigr]
                 \psi - \frac{{C_0}}{2}(\psi^\dagger \psi)^2
                    - \frac{{D_0}}{6}(\psi^\dagger \psi)^3 +  \ldots
 \eeq
where the parameters can be determined from
free-space scattering (or by finite density fits); dimensional
regularization with minimal subtraction is particularly advantageous.
The EFT for the uniform system (no external potential) predicts
the energy (and other observables) 
simply and efficiently through a diagrammatic expansion,
with a power counting that tells us what diagram to calculate at
each order in the expansion.
[See \cite{HAMMER00} for details.]
The first terms in the diagrammatic expansion are
(the solid dot is $C_0 = 4\pi a_s/{M}$, with $a_s$ the scattering 
length): 
\begin{figure}[h]
  \centering
  \includegraphics*[width=3.1in,angle=0.0]{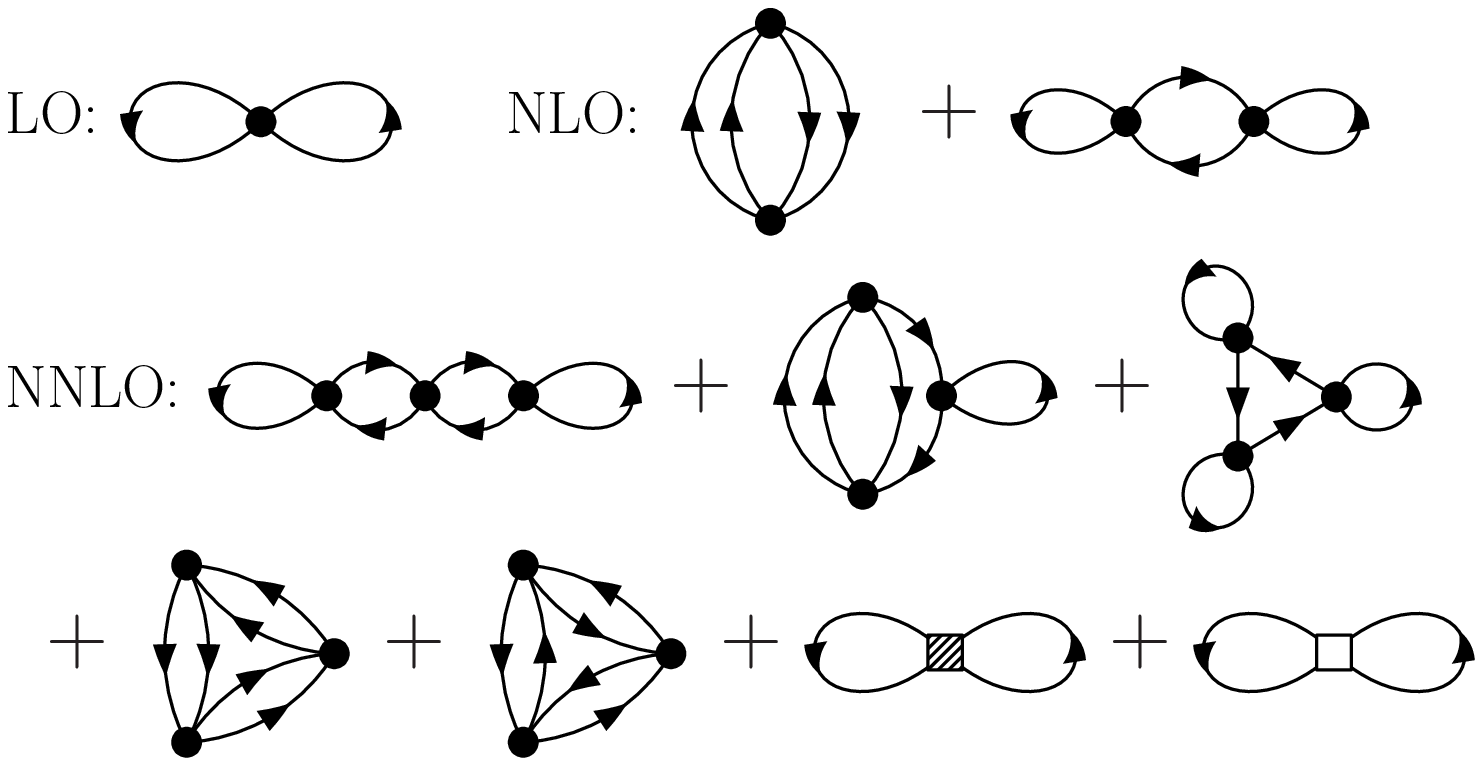}
%  \caption{Diagrams.}
  \label{fig:fig_hug8}
\end{figure}

\noindent
which yields the expansion for the energy density (with degeneracy $g$):
\beq
    {\cal E} =
       n \frac{{\kf^2}}{2M}
        \biggl[ \frac{3}{5} + (g-1)\biggl\{\frac{2}{3\pi}{(\kf a_s)}
        +
          \frac{4}{35\pi^2}(11-2\ln 2){(\kf a_s)^2}
           \biggr\} + \cdots   \biggr]    
\eeq      
in powers of $\kf a_s$.  (Note this is \emph{not} a power series;
see \cite{HAMMER00}.)
The point here is that we have a systematic expansion: LO, NLO, NNLO,
and so on, just as in the chiral NN case.
To put the gas in a harmonic or gaussian trap with external
potential $v({\bf x})$, we simply add a
$v({\bf x})\rho({\bf x})$ term to the Lagrangian.

The effective action formalism starts 
with a path integral for the generating functional with an
additional source $J$ coupled to the density:
\beq
    Z[J] = e^{i W[J]} =
    \int\! {\cal D}\psi^\dagger {\cal D}\psi\
      e^{i\int\!d^4x\, [{\cal L} - {J(x)\psi^\dagger(x)\psi(x)}]}
      \ .
\eeq
(Different external sources can be used.)
The density is found by a functional derivative with respect to
$J$:
\beq
   \rho(x) \equiv \langle \psi^\dagger(x)\psi(x) \rangle_J
      = \frac{\delta W[J]}{\delta J(x)} \ .        
  \label{eq:rhoJ}
\eeq
The crucial step is the inversion of (\ref{eq:rhoJ}); that is,
to find $J[\rho]$.  Given that, the effective action
$\Gamma[\rho]$ follows from a functional Legendre transformation:
\beq
  \Gamma[\rho] = W[J] - \int\!d^4x\, J(x) \rho(x) \ .
\eeq
The functional $\Gamma[\rho]$ is proportional (with a trivial time
factor) to the ground-state energy functional $E[\rho]$ that we seek.
The density $\rho(x)$ is determined by stationarity:
\beq
  \frac{\delta\Gamma[\rho]}{\delta \rho(x)} = - J(x) \Longrightarrow 0
      \mbox{\ \ in the ground state.} 
      \label{eq:legendre}
\eeq 
That's density functional theory in a nutshell!

Two approaches to invert the Legendre transformation have been proposed.
They correspond to generating point coupling or meson-based models.
We'll outline the first approach, 
which is the inversion method of Fukuda et al.
\cite{Fukuda:im,VALIEV97}. It is a generalization of
the Kohn-Luttinger inversion procedure \cite{KOHN60}.
The idea is to use an order-by-order matching in a counting parameter
$\lambda$, which we identify with an EFT expansion parameter
(e.g., $1/\Lambda_\chi$).  We expand each of the ingredients of
(\ref{eq:legendre}):
\beqa
   J[\rho,\lambda] &\!=\!& J_0[\rho] + \lambda J_1[\rho] 
       + \lambda^2 J_2[\rho] + \cdots \\
   W[J,\lambda] &\!=\!& W_0[J] + \lambda W_1[J] 
       + \lambda^2 W_2[J] + \cdots \\
   \Gamma[\rho,\lambda] &\!=\!& \Gamma_0[\rho] 
            + \lambda \Gamma_1[\rho] 
            + \lambda^2 \Gamma_2[\rho] + \cdots 
\eeqa
and equate each order in $\lambda$.
Zeroth order corresponds to a noninteracting system with
potential $J_0(x)$:
\beq
  \Gamma_0[\rho] = W_0[J_0] - \int\!d^4x\, J_0(x)\rho(x) 
   \quad \Longrightarrow \quad \rho(x) 
   = \frac{\delta W_0[J_0]}{\delta J_0(x)} \ .
\eeq
This is the Kohn-Sham system with the \emph{exact} density!
That is, $J_0$ is the potential that reproduces the exact interacting
density in a non-interacting system.
We diagonalize $W_0[J_0]$ by introducing Kohn-Sham orbitals,
which yields a sum of eigenvalues.
Finally, we find $J_0$ for the ground state by completing a
self-consistency loop:
 \beq
   J_0 \rightarrow W_1 \rightarrow \Gamma_1 \rightarrow J_1
    \rightarrow W_2 \rightarrow \Gamma_2 \rightarrow \cdots
    \Longrightarrow
      J_0(x) = \sum_i \frac{\delta\Gamma_{i}[\rho]}{\delta\rho(x)}
      \ .    
 \eeq

We can contrast the conventional  expansion of the
propagator:
\begin{center}
\includegraphics*[angle=0.0,width=2.35in]{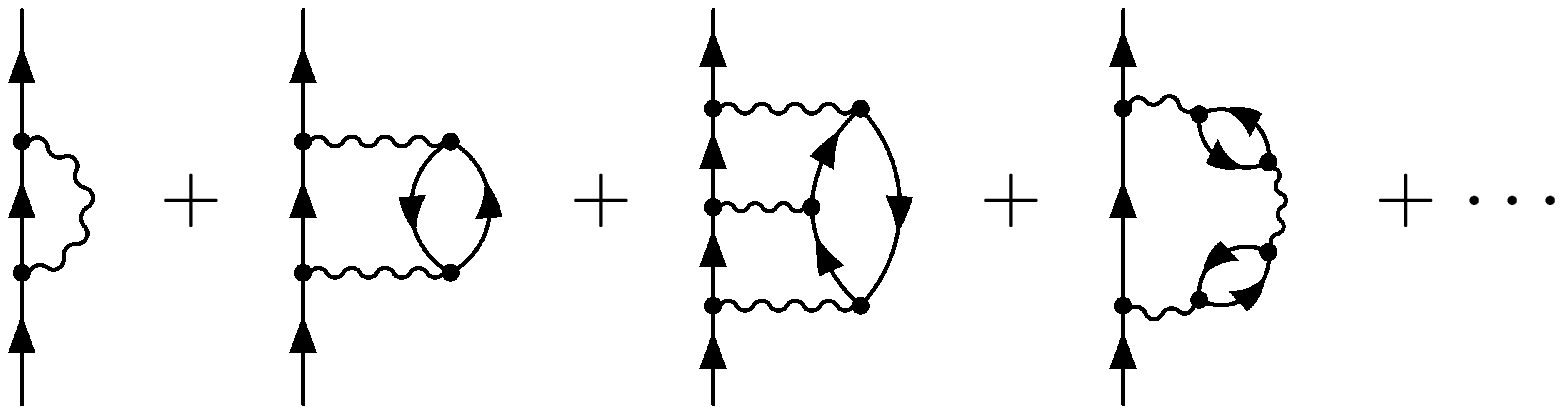}
    \raisebox{.29in}{$\Longrightarrow\ $}
\includegraphics*[angle=0.0,width=1.9in]{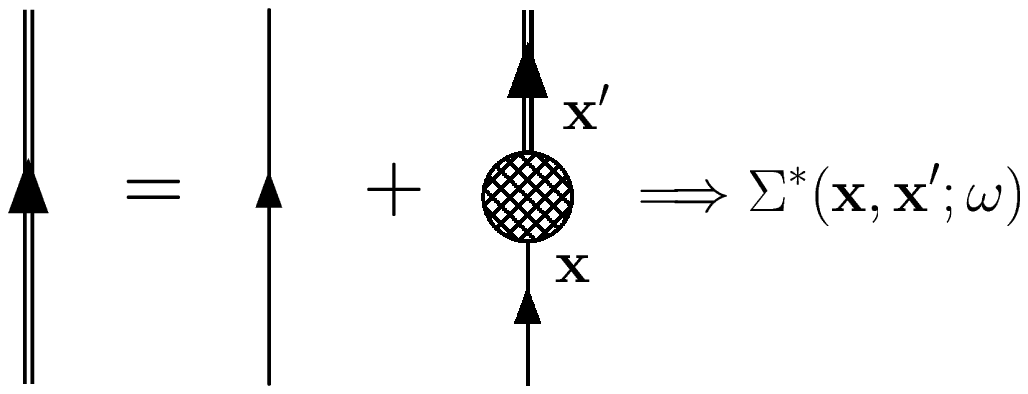}
\end{center}
which involves a non-local, state dependent
$\Sigma^\ast({\bf x},{\bf x}';\omega)$,
with the local $J_0({\bf x})$:
\begin{center}
  \includegraphics*[angle=0.0,width=3.2in]{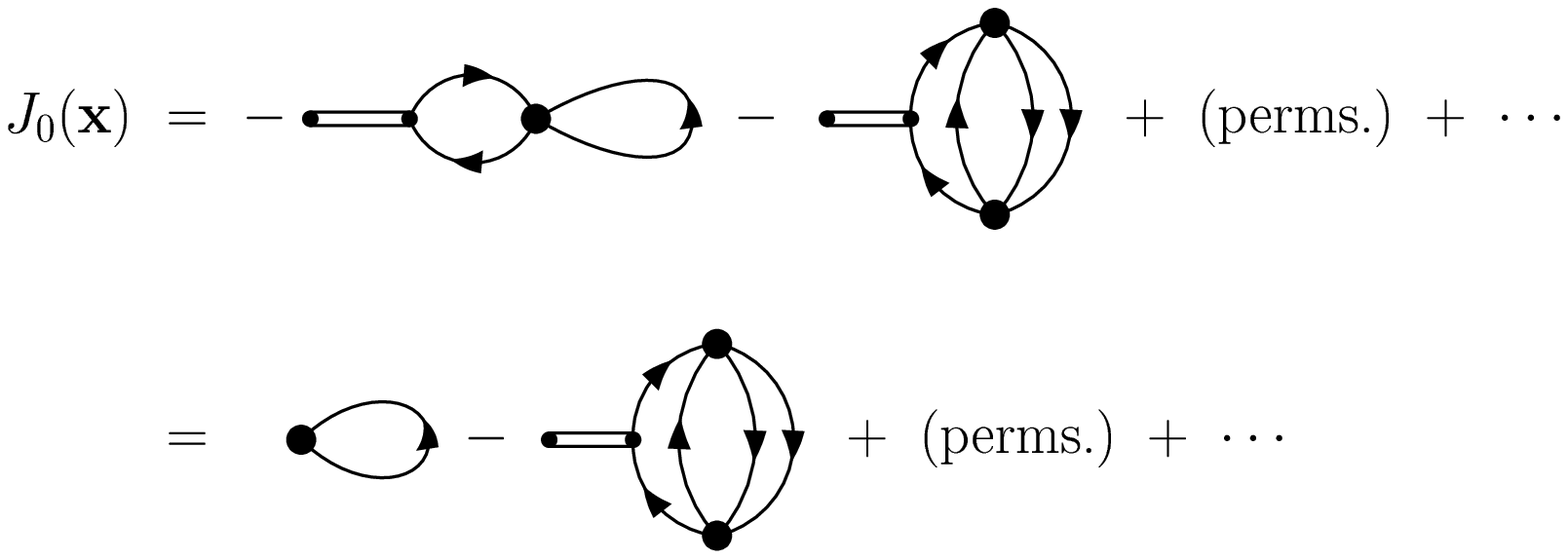}
\end{center}
To carry out this construction there are new Feynman rules that involve
this ``inverse density-density correlator,'' 
which is indicated by a double line:
\begin{center}
  \includegraphics*[angle=0.0,width=4.2in]{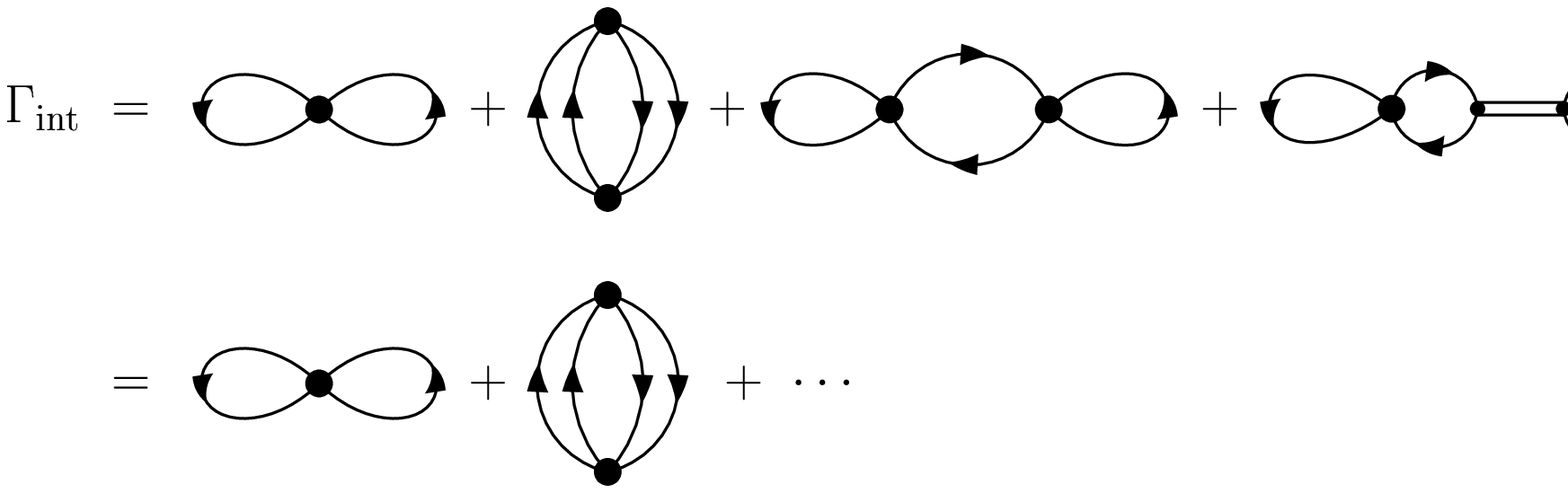}
\end{center}
Complete details are available in \cite{PUGLIA03}.

\begin{figure}[t]
  \centering
  \includegraphics*[width=3.9in,angle=0.0]{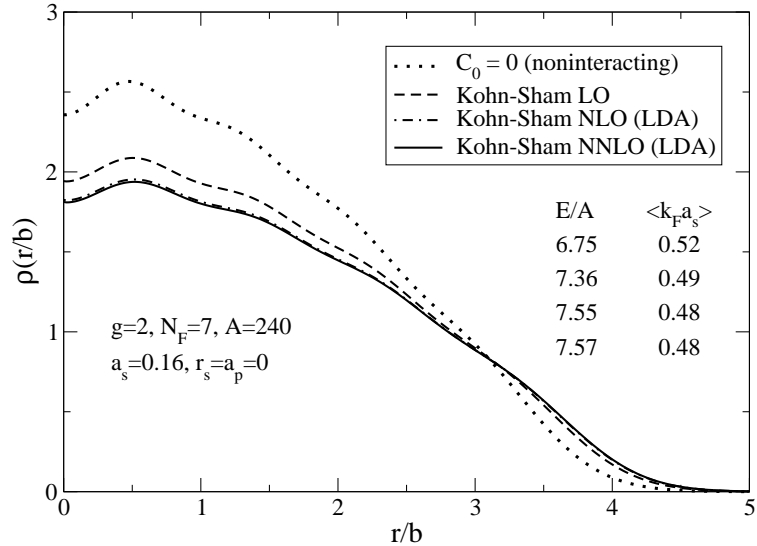}  
  \caption{Density profiles for a dilute Fermi gas in 
  a harmonic trap for successive approximations.}
  \label{fig:dft1}
\end{figure}

If we restrict the application to closed shells for simplicity,
we just specify the spin degeneracy $g$ and the Fermi harmonic oscillator
shell (the last occupied shell) $N_F$, which then determines the number
$A$ of atoms.
The iteration procedure for the dilute Fermi gas in a harmonic
trap is as follows:
\begin{enumerate}
   \item guess an initial density profile $\rho(r)$
      (e.g., Thomas-Fermi),
   \item evaluate the local single-particle potential 
    %\mtp{v_{\rm eff}[\rho(r)] \equiv
      $v_{\rm eff}(r) \equiv v(r) - J_0(r)$,
   \item solve for the lowest $A$ states (including degeneracies) 
        $\{\psi_\alpha,\epsilon_\alpha\}$:
    \beq
     \bigl[ -\frac{{\nabla}^2}{2M}  +  v_{\rm eff}(r)
     \bigr]\, \psi_\alpha({\bf x}) = 
     \epsilon_\alpha \psi_\alpha({\bf x}) \ ,
    \eeq
   \item compute a new density
          $\rho(r) 
            = \sum_{\alpha=1}^{A} |\psi_\alpha({\bf x})|^2$;
       other observables are functionals of 
          $\{\psi_\alpha,\epsilon_\alpha\}$,
   \item repeat 2.--4. until changes are small (``self-consistent'').
\end{enumerate}
This procedure is simple to implement as a computer program and
will be familiar to anyone who has performed a relativistic
mean-field calculation for finite nuclei.

\begin{figure}
  \centering
  \includegraphics*[width=3.9in,angle=0]{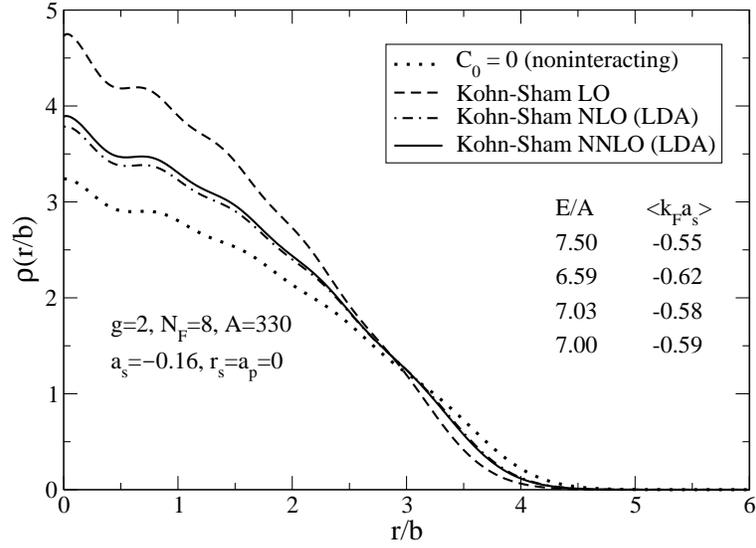}  
  \caption{Density profiles for a dilute Fermi gas in 
  a harmonic trap for successive approximations.}
  \label{fig:dft2}
\end{figure}

Some representative results are given in Figs.~\ref{fig:dft1} and
\ref{fig:dft2}.
Note the convergence of the densities and energies
as we increase the order of the calculation.
In Fig.~\ref{fig:dftest}, we show the contribution to the energy
per particle at LO, NLO, and NNLO for three different systems.
We estimate the expected contribution by NDA from the Hartree terms in
the functional (which appear with simple powers of the density) and then
scale correction terms by the expansion parameter $\langle \kf \rangle
a_s$.  The power counting still works!  This is an excellent laboratory
to explore such issues.  For example, the rightmost circle shows a large
discrepancy between predicted and actual values.  A closer examination
reveals an ``accidental'' cancellation (that goes away for different
parameter choices!).

\begin{figure}
  \centering
  \includegraphics*[width=2.5in,angle=0.0]{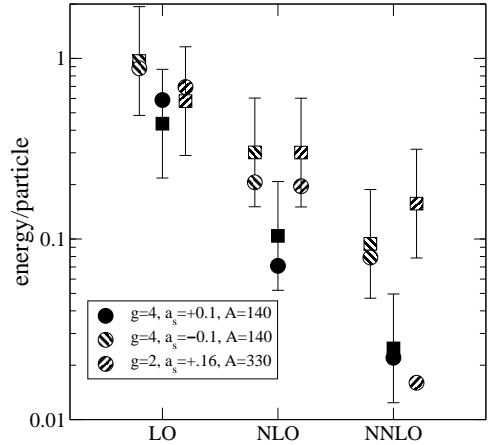}
  \caption{Energy estimates (squares) 
  and actual values (circles) for a dilute system.}
  \label{fig:dftest}
\end{figure}

Although we did not discuss covariant DFT here, Schmid et al.\ have
already laid the foundation \cite{SCHMID95}.  
The generalization of DFT to other sources and time
dependence is straightforward (cf.\ spin-density functionals for Coulomb
systems).  For the covariant case, one treats $\overline\psi_i\psi_j$ as
a matrix and introduces a source $j_{ij}$ coupled to it, which yields a
functional of the scalar density and vector current, as described
in Sect.~3.3.

%%%%%%%%%%%%%%%%%%%%%%%%%%%%%%%%%%%%%%%%%%%%%%%%%%%%%%%%%%%%%%%%%%%%%
%%%%%%%%%%%%%%%%%%%%%%%%%%%%%%%%%%%%%%%%%%%%%%%%%%%%%%%%%%%%%%%%%%%%%

\section{Building the Next Generation \emph{Theories}}

The successes of chiral EFT for nucleon-nucleon scattering and the
few-body problem have raised the bar for models of nuclear structure.
For the next generation,
we must justify what we do and not just mumble ``effective theory'' to
account for phenomenological successes.  This requires that we develop
and test systematic power counting schemes and reexamine
and sharpen arguments for a covariant approach.
To meet the challenge, the next generation of covariant models must move
away from model dependence.
This implies turning away from the ``minimal model'' philosophy and
including all terms allowed by symmetries (up to redundancies).
It means requiring error estimates and a controlled expansion in density,
asymmetry, and momentum transfer.  And it means that the results should
exhibit regulator (e.g., cutoff) independence up to expected
truncation errors.

There has been significant progress made toward building the next
generation of covariant approaches to nuclear structure (e.g., see B.D.\
Serot's lecture).
But there is a long list of loose ends to address.
Some of the physics goals and open issues are 
\begin{itemize}
  \item a reliable description of nuclei far from stability;
  \item controlled equation of state and transport properties
  of neutron stars;
  \item unified description of collective excitations;
  \item currents, both elastic and transition currents to 
    excited states;
  \item improved treatment of isovector physics;
  \item consistent incorporation of chiral physics;
  \item consistent treatment of pairing in covariant density
     functionals;
  \item establishing more direct 
    connections to QCD (e.g., quark mass dependence of
    equilibrium properties).
\end{itemize}
The framework we have proposed to address these goals and issues is a merger of
effective field theory and density functional theory.  The idea is to
embed the successful relativistic mean-field phenomenology into this
framework, as traditional NN interactions are taken over by chiral EFT.
The basic structure and successes are preserved, but model dependence is
removed.

There are many challenges.
For the covariant many-body EFT, one needs a consistent application of
infrared regularization (or equivalent) applied at finite density.
This is likely to mean the use of cutoff regularization.
A possible laboratory for these issues is a dilute Fermi gas with 
$\Lambda \sim M_N$, where in a finite system the spin-orbit force can be
directly explored.
In addition,
chiral EFT expansions must be translated to covariant form.
For the covariant DFT, one needs systematic derivative expansions
(analogous to past density matrix expansions), the incorporation of
long-range forces and pairing, and a consistent extension to
time-dependent DFT. 
Most importantly, we need a more complete power counting scheme for
energy functionals.

There is growing
hope for an ultimate path to \emph{ab initio} calculations of
nuclei, starting from quantum chromodynamics (QCD).
Nonperturbative lattice QCD could calculate low-energy constants needed
for the chiral EFT of NN and few-body systems.  This EFT in turn
determines the analytic structure and low-energy constants of a
(covariant) many-body EFT, which takes the form of a DFT energy
functional, applicable across the periodic table.
There are many gaps, but this dream has become significantly 
more plausible at all levels in just the last few years.

%%%%%%%%%%%%%%%%%%%%%%%%%%%%%%%%%%%%%%%%%%%%%%%%%%%%%%%%%%%%%%%%%%%%%
%%%%%%%%%%%%%%%%%%%%%%%%%%%%%%%%%%%%%%%%%%%%%%%%%%%%%%%%%%%%%%%%%%%%%

%%%%%%%%%%%%%%%%%%%%%%%%%%%%%%%%%%%%%%%%%%%%%%%%%%%%%%%%%%%%%%%%%%%%%
%%%%%%%%%%%%%%%%%%%%%%%%%%%%%%%%%%%%%%%%%%%%%%%%%%%%%%%%%%%%%%%%%%%%%
% Non-BibTeX users please follow the syntax
% the syntax of "references.tex" for your own citations
%%%%%%%%%%%%%%%%%%%%%%%% referenc.tex %%%%%%%%%%%%%%%%%%%%%%%%%%%%%%
% sample references
% "physics"
%
% Use this file as a template for your own input.
%
%%%%%%%%%%%%%%%%%%%%%%%% Springer-Verlag %%%%%%%%%%%%%%%%%%%%%%%%%%

%
% BibTeX users please use
% \bibliographystyle{}
% \bibliography{}

%%%%%%%%%%%%%%%%%%%%%%%%%%%%%%%%%%%%%%%%%%%%%%%%%%%%%%%%%%%%%%%%%%%%%%  }

%%%%%%%%%%%%%%%%%%%%%%%%%%%%%%%%%%%%%%%%%%%%%%%%%%%%%%%%%%%%%%%%%%%%%%

\printindex
\end{document}